\documentclass[aps,a4paper,11pt,tightenlines,titlepage]{revtex4}%
\usepackage{amsfonts}
\usepackage{amsmath}
\usepackage{amssymb}
\usepackage{graphicx}%
\begin{document}
\title{Optical M\"{o}bius Strips in Three Dimensional Ellipse Fields: Lines of
Circular Polarization }
\author{Isaac Freund}
\affiliation{Department of Physics, and Jack and Pearl Resnick Advanced Technology
Institute, Bar-Ilan University, Ramat-Gan 52900, Israel}

\begin{abstract}
The major and minor axes of the polarization ellipses that surround singular
lines of circular polarization in three dimensional optical ellipse fields are
shown to be organized into M\"{o}bius strips. \ These strips can have either
one or three half-twists, and can be either right- or left-handed. \ The
normals to the surrounding ellipses generate cone-like structures. Two special
projections, one new geometrical, and seven new topological indices are
developed to characterize the rather complex structures of the M\"{o}bius
strips and cones. \ These eight indices, together with the two well-known
indices used until now to characterize singular lines of circular
polarization, could, if independent, generate $16,384$ geometrically and
topologically distinct lines. \ Geometric constraints and $13$ selection rules
are discussed that reduce the number of lines to $2,104$, some $1,150$ of
which have been observed in practice; this number of different C lines is
$\sim350$ times greater than the three types of lines recognized previously.
\ Statistical probabilities are presented for the most important index
combinations in random fields. \ It is argued that it is presently feasible to perform
experimental measurements of the M\"{o}bius strips and cones described here
theoretically.

\end{abstract}

\maketitle

\section{INTRODUCTION}

Three dimensional (3D) optical fields are, except in special cases,
elliptically polarized. \ In paraxial fields the polarization ellipses lie in
parallel planes oriented normal to the propagation direction, but in 3D fields
the ellipses generally have a wide range of spatial orientations.

\begin{figure}
[h]
\includegraphics[width=0.8\textwidth]%
{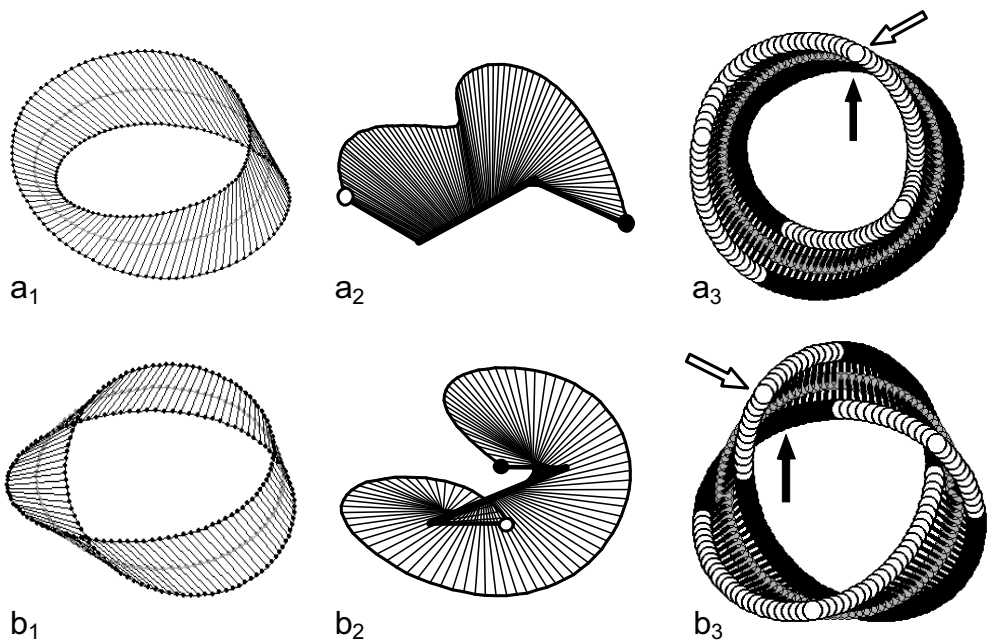}%
\caption{M\"{o}bius strips surrounding C lines in a computed random optical
field. \ The strips shown here are generated by the major axes of the ellipses
located on a small circle that encircles the line. \ The M\"{o}bius strips in
(a$_{1}$)$-$(a$_{3}$) contain a single half-twist, those in (b$_{1}$%
)$-$(b$_{3}$) three half-twists.\ \ (a$_{1}$) and (b$_{1}$)$\ $ are overall
views of the strips that show their 3D structure. \ Here ellipse centers are
shown by small gray dots, the major axes of the ellipses by thin straight
lines, and the endpoints of these axes by small black dots. \ (a$_{2}$) and
(b$_{2}$) show the circular strips in (a$_{1}$) and (b$_{1}$)$\ $cut open and
straightened to better display the twist structure. \ Here the ellipse centers
form the thick central line, and, for clarity, only one half of the ellipse
axes are shown (thin straight lines). \ The small white (black) circles mark
the endpoints of the first (last) ellipse, and serve to emphasize the twist
structure; when the strips are closed these ellipses are adjacent, but their
endpoints lie on opposite sides of the circle of ellipse centers. \ (a$_{3}$)
and (b$_{3}$) are views from above that illustrate the fact that these strips,
which have an odd number of half-twists, have not only one side, but also have
only one edge. \ Here the centers of the ellipses are shown by gray circles,
the endpoints of ellipse axes that lie above (below) the plane of ellipse
centers by white (black) circles. \ Starting at the highest point on the
\textquotedblleft top\textquotedblright\ edge (white arrow) and traversing a
$360^{o}$ circuit along this edge one arrives at the \textquotedblleft
bottom\textquotedblright\ edge (black arrow). \ Continuing in the same
direction along a second $360^{o}$ circuit one returns to one's starting point
on the \textquotedblleft top\textquotedblright\ edge, showing that there is,
in fact, only a single edge to the strip. \ The ellipse endpoints in (b$_{3}$)
can be seen to form a trefoil knot, whereas those in (a$_{3}$) do not form a
knot. }%
\label{Fig1}%
\end{figure}

The generic singularities of 3D ellipse fields are meandering lines of
circular polarization, C lines, and meandering lines of linear polarization, L
lines [$1-16$]. \ Here we study the arrangement of the polarization ellipses
surrounding C lines; the arrangements of the polarization ellipses that
surround L lines will be reported on separately. \ Where a C line pierces a
plane, $\Sigma$, a point of circular polarization, a C point, appears.
\ Previous studies of C lines have concentrated on the two dimensional
projections onto $\Sigma$ of those ellipses whose centers lie in this plane on
a circle that encloses the C point. \ Examining the \emph{full} 3D arrangement
of these ellipses, we find that their major and minor axes generate M\"{o}bius
strips with either one or three half-twists; examples of such strips are shown
in Fig. \ref{Fig1}.%

\begin{figure}
[h]
\includegraphics[width=0.90\textwidth]%
{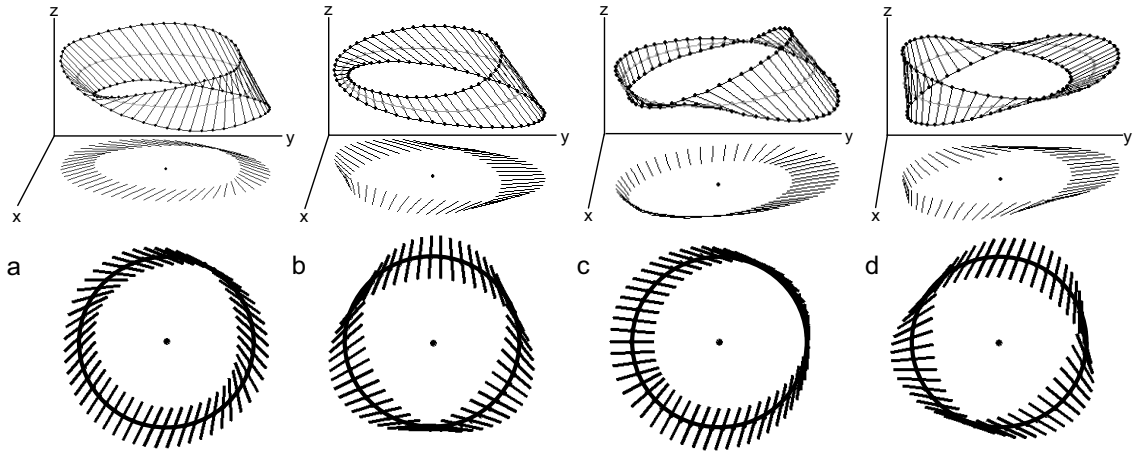}%
\caption{C point winding number $I$. \ Shown in panels (a) and (b) ((c) and
(d)) are C point M\"{o}bius strips with one (three) half-twists, upper panels,
and their projections onto the $xy$-plane, lower panels. \ As in Figs. 1a and
1b, in the upper panels ellipse centers are shown by small gray dots, major
axes of the ellipses are shown by thin straight lines, and the endpoints of
these axes by small black dots. \ The central black dot is the C point. \ In
the lower panels the ellipse centers form the black circle, and the
projections of the ellipse axes are shown by short straight lines. During one
$360^{o}$ circuit around the C point in say the counterclockwise direction, in
(a) and (c) ((b) and (d)) these lines rotate by $180^{o}$ in the same
counterclockwise (opposite, clockwise) direction as the circuit, and $I=+1/2$
($-1/2$).}%
\label{Fig2}%
\end{figure}

The three orthogonal principal axes of the (always planar, [$17$, Sect. 1.4])
polarization ellipse are its major and minor axes, and the normal to the
ellipse. \ On a C line the major and minor axes of the ellipse become equal,
and the ellipse degenerates into a circle, the C circle. \ Because a circle
has no preferred direction, its major and minor axes are undefined (singular).
\ The normal to the C circle, however, remains well defined, as do all three
principal axes of the ellipses that surround the C line. \ The projections of
the major and minor axes of the ellipses whose centers lie in a plane $\Sigma$
pierced by a C line rotate about the C point, generically with winding number
(net rotation or winding angle divided by $2\pi$) $I=$ $\pm1/2$ [$1,3,10$].
\ Examples of such rotations are shown in Fig. \ref{Fig2} for the major axis;
in all known cases, for a given C point the winding number of the projections
of the minor axes onto $\Sigma$\ is the same as that of the major axes
[$1,3,10$]%

Winding number $I$ and a geometrical index described later, the line index
[$1,3,18$], fully characterize the projections in Fig. \ref{Fig2}; together
these indices lead to the currently known three distinct C lines [$1,3$]. \ As
is evident from Fig. \ref{Fig2}, however, these projections of the major/minor
axes are insufficient to determine the properties, or even the existence, of
the parent M\"{o}bius strips. \ In later sections we introduce one new
geometrical, and seven new topological indices to characterize these strips;
together these eight new indices increase the number of distinct C lines by a
factor of $\sim350$.

The plan of this report is as follows: \ In Section II we describe the
analytical and numerical tools used in the computer simulations employed in
later sections to study 3D ellipse fields, their C lines, their M\"{o}bius
strips, and related structures. \ These tools are similar to those we used to
study the one-full-twist M\"{o}bius strips that surround ordinary (i.e.
nonsingular) ellipses [$14,15$]. In Section III we describe in detail the
M\"{o}bius strips and related structures that surround C lines, introduce the
new indices that characterize these lines, and extend the important line
classification of C points [$1,3,18$]; although this classification does not
involve an invariant topological index, it does serve to further characterize
the arrangement of the ellipses that surround these points. \ In Section IV we
present statistical data for the relative occurrences in random fields of the
many different C lines. \ We summarize our main findings in the concluding
Section V.

Coherent measurement techniques permit the determination (amplitude and phase)
of all three orthogonally polarized components of 3D microwave fields
[$6,7,19,20$]; recent advances in interferometric nanoprobes provide similar
capabilities for optical fields [$21-28$]. \ It is therefore now possible to
carry out experiments that can measure the highly unusual structures described
here. \ C lines are degeneracies of the matrix that describes the 3D
polarization ellipse [$9,29,30$]. \ Other physical systems such as liquid
crystals, strain fields, flow fields, etc., are described by similar matrices
whose degeneracies can be expected to yield analogs of C lines that are likely
to be surrounded by M\"{o}bius strips.

At present there are no known practical applications for optical M\"{o}bius
strips; as these strange, engrossing objects become better understood,
however, useful applications may emerge.

\section{METHODS}

As indicated in the Introduction, the only well defined direction associated
with a C point is the normal to the C circle. \ We call the plane that is
perpendicular to this preferred direction, i.e. the plane of the C circle, the
\emph{principal plane}, and denote this plane by $\Sigma_{0}$. \ We emphasize
that $\Sigma_{0}$\ need not be, and in general is not, perpendicular to the C
line, and that the orientation of the principal plane relative to the C line
changes as one moves along the line [$1,3,8-10,13-15$].

\subsection{Ellipse Axes}

Here we describe the methods we use to calculate the axes of the polarization
ellipses that generate the M\"{o}bius strips surrounding C points and C lines.

We label the major and minor axes of, and the normal to, the general
polarization ellipse by \textbf{$\boldsymbol{\alpha}$},
\textbf{$\boldsymbol{\beta}$}, and \textbf{$\boldsymbol{\gamma}$},
respectively. \ Given an expression for the (here complex) optical field
$\mathbf{E}$, \textbf{$\boldsymbol{\alpha}$}, \textbf{$\boldsymbol{\beta}$},
and \textbf{$\boldsymbol{\gamma}$}, can be calculated in two seemingly
different ways. \ The first, due to Berry [$8,10$], is
\begin{subequations}
\label{alphabetagamma}%
\begin{align}
\boldsymbol{\alpha}  &  =\operatorname{Re}(\mathbf{E}^{\ast}\sqrt
{\mathbf{E}\boldsymbol{\cdot}\mathbf{E}}),\label{alpha}\\
\boldsymbol{\beta}  &  =\operatorname{Im}(\mathbf{E}^{\ast}\sqrt
{\mathbf{E}\boldsymbol{\cdot}\mathbf{E}}),\label{beta}\\
\boldsymbol{\gamma}  &  =\operatorname{Im}\left(  \mathbf{E}^{\ast
}\mathbf{\times E}\right)  . \label{gamma}%
\end{align}
The second involves finding the three eigenvalues, $\lambda_{i}$, and three
normalized eigenvectors, $\boldsymbol{\nu}_{i}$, $i=1,2,3$, of the $3\times3$
real coherency matrix,
\end{subequations}
\begin{equation}
M_{ij}=\operatorname{Re}(E_{i}^{\ast}E_{j});\;i,j=x,y,z. \label{RealCohMatrix}%
\end{equation}
The largest eigenvalue, here $\lambda_{3}$, is associated with the major axis
of the polarization ellipse, the next largest eigenvalue, $\lambda_{2}$, with
the minor axis, and the smallest eigenvalue $\lambda_{1}\equiv0$, with the
normal to the ellipse.

$\lambda_{1}\equiv0$ because for the monochromatic fields assumed here the
polarization ellipse is always planar [$17$]. \ In terms of $\lambda$, the
length $a$ of the major axis of the polarization ellipse is given by
$a=\sqrt{\lambda_{3}}$, the length $b$ of the minor axis by $b=\sqrt
{\lambda_{2}}$.

The above two seemingly different methods are reconciled by noting that
$\boldsymbol{\alpha}$$\boldsymbol{/}\left\vert \boldsymbol{\alpha}\right\vert
=\boldsymbol{\nu}_{3}$, $\boldsymbol{\beta}$$\boldsymbol{/}\left\vert
\boldsymbol{\beta}\right\vert =\boldsymbol{\nu}_{2}$, and $\boldsymbol{\gamma
}$$\boldsymbol{/}\left\vert \boldsymbol{\gamma}\right\vert =\boldsymbol{\nu
}_{1}$.\ \ Although an analytical proof of these equalities is still lacking,
we note that in every one of the literally hundreds of cases studied these
equalities were found to hold to within numerical accuracy. \ In what follows,
\textbf{$\boldsymbol{\alpha}$}, \textbf{$\boldsymbol{\beta}$}, and
\textbf{$\boldsymbol{\gamma}$} are, without change of notation, understood to
each be normalized to unit length.

As we move through the wavefield, in order to ensure that
\textbf{$\boldsymbol{\alpha}$} and \textbf{$\boldsymbol{\beta}$} are smooth,
continuous, single valued functions of position, we calculate $\sqrt
{\mathbf{E}\boldsymbol{\cdot}\mathbf{E}}$ as follows: \ We write
$\mathbf{E}\boldsymbol{\cdot}\mathbf{E=}E_{x}^{2}+E_{y}^{2}+E_{z}^{2}%
=A\exp\left(  \text{i}\mathbb{\varphi}\right)  $, unfold (unwrap)
$\mathbb{\varphi}$ as needed to eliminate spurious discontinuities of $2\pi$,
and then write $\sqrt{\mathbf{E}\boldsymbol{\cdot}\mathbf{E}}\mathbf{=}%
\sqrt{A}\exp\left(  \text{i}\mathbb{\varphi}/2\right)  $.

Although the polarization ellipse has inversion symmetry, $\boldsymbol{\gamma
}$ in Eq. (\ref{gamma}) is an axial vector that defines a unique positive
direction. \ This direction, which is determined via a right hand rule from
the way $\mathbf{E}$ traces out the polarization ellipse in time, permits us
to define a unique orthogonal $xyz$ coordinate system in which the positive
$z$-axis is along the positive direction of $\boldsymbol{\gamma}$, the $x$ and
$y$ axes are along \textbf{$\boldsymbol{\alpha}$} and
\textbf{$\boldsymbol{\beta}$}, and $xyz$ form a right-handed $3$-frame.

C lines may be located in two different ways: \ The first is to locate the
zeros of $\mathbf{E}\boldsymbol{\cdot}\mathbf{E}$ [$8-10$]. \ The second is to
locate the zeros of the C point discriminant $D_{C}=a_{1}^{2}-4a_{2}$; this
discriminant is obtained from the characteristic equation $\lambda^{2}%
+a_{1}\lambda+a_{2}=0$ of the real coherency matrix $M_{ij}$ in Eq.
(\ref{RealCohMatrix}) [$29$]. \ As expected, both methods were found to yield
the same result.

\subsection{Axis Projections}

Having traced out a C line, we move along the line to some C point and find
$\boldsymbol{\gamma}$ for that point. \ As noted above, $\boldsymbol{\gamma}$
for a given point is the positive normal to $\Sigma_{0}$ at the point. \ In
looking at the ellipses in $\Sigma_{0}$ we will always look \emph{down} the
positive $z$-axis (i.e. $+z$ points \emph{towards} us).

In constructing the M\"{o}bius strips and their projections we focus on those
ellipses whose centers lie in $\Sigma_{0}$ on a small circle that surrounds
the C point. \ We call these ellipses the \emph{surrounding ellipses} and
label the corresponding circle $\sigma_{0}$. \ For convenience, we take the
center of $\sigma_{0}$ to coincide with the C point, although any other simple
path in $\Sigma_{0}$ centered on the C point yields the same result for the
various topological indices as does $\sigma_{0}$.

\begin{figure}
[h]
\includegraphics[width=0.85\textwidth]%
{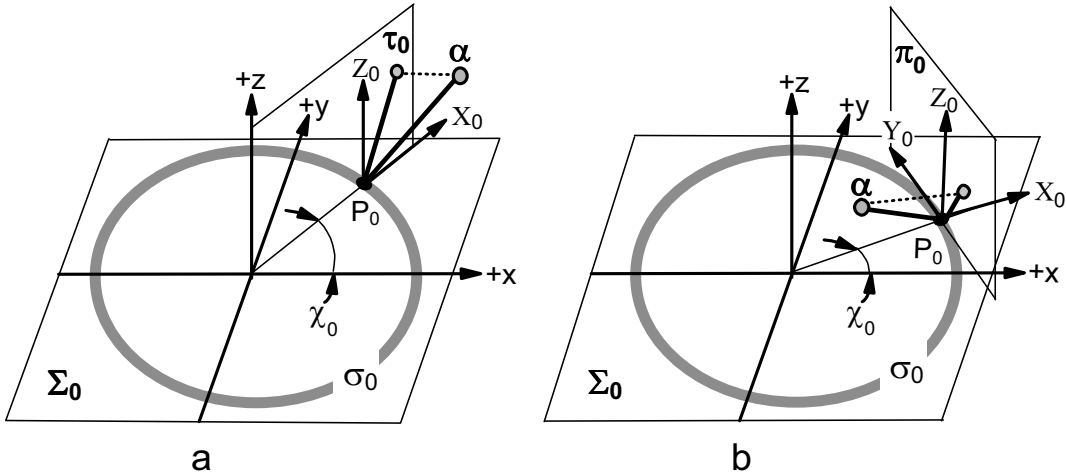}%
\caption{Projections $\tau_{0}$ and $\pi_{0}$. \ Shown as an example is axis
$\boldsymbol{\alpha}$ projected onto plane $\tau_{0}$ (plane $\pi_{0}$) in (a)
in ((b)). \ Other symbols are explained in the text in Section IIB. \ Position
on $\sigma_{0}$ is parameterized by the angle $\chi_{0}$, or by arclength
$s_{0}=r_{0}\chi_{0}$, where $r_{0}$ is the radius of $\sigma_{0}$.}%
\label{Fig3}%
\end{figure}

In addition to projecting the ellipse axes onto $\Sigma_{0}$, we study two
other projections. \ The first, and most important, we call the $\tau_{0}$
projection. \ In this projection we erect a rotating plane, the $\tau_{0}$
plane. \ $\tau_{0}$ is oriented normal to $\Sigma_{0}$, contains the normal to
the C circle, axis $\boldsymbol{\gamma}$ (the $+z$ axis), and rotates around
this axis. \ Where $\sigma_{0}$ pierces the plane we establish an orthogonal
$X_{0}Z_{0}$-coordinate system. \ The angle of rotation $\chi_{0}$ of the
plane $\tau_{0}$ is measured counterclockwise from the fixed (laboratory)
$x$-axis in $\Sigma_{0}$. \ Projected onto $\tau_{0}$ is the axis of the
ellipse whose center lies in $\sigma_{0}$ at the point where $\tau_{0}$
intercepts this circle; we label this point $P_{0}$. $P_{0}$, of course, moves
along $\sigma_{0}$ as $\chi_{0}$ increases $0-2\pi$. \ Fig. \ref{Fig3}a
displays the relevant geometry.

The second projection we call the $\pi_{0}$ projection. \ In this projection
we erect another rotating plane, the $\pi_{0}$ plane. \ Like $\tau_{0}$,
$\pi_{0}$ is oriented normal to $\Sigma_{0}$, but unlike $\tau_{0}$, $\pi_{0}$
is \emph{tangent} to $\sigma_{0}$. \ We establish an orthogonal $Y_{0}Z_{0}%
$-coordinate system at the point of tangentsy, the point $P_{0}$, such that
the triplet $X_{0}Y_{0}Z_{0}$ forms a right handed coordinate system. \ The
same angle, $\chi_{0}$, that measures the rotation of $\tau_{0}$ measures the
rotation of $\pi_{0}$. \ Projected onto $\pi_{0}$ is the axis of the ellipse
whose center lies at $P_{0}$. \ Fig. \ref{Fig3}b illustrates the geometry of
this case.

Thus, in determining the geometrical and topological properties of the 3D
M\"{o}bius strips formed by a given axis we use three orthogonal projections
$-$ the minimum number of projections required for the characterization of a
three-dimensional object.

In general, there are therefore nine different projections: the projections
onto the three planes $\Sigma$, $\tau$, and $\pi$, of each of the three axes
$\boldsymbol{\alpha}$, $\boldsymbol{\beta}$, and $\boldsymbol{\gamma}$ of the
ellipses on $\sigma$. \ Although these projections are interconnected, as
becomes apparent, they yield a multitude of different topological winding
numbers \ This may appear surprising because, after all, the three-frame
$\boldsymbol{\alpha\beta\gamma}$ has only three rotational degrees of freedom.
\ But the properties of the M\"{o}bius strips depend on the rotations of
\emph{all} the ellipses on $\sigma_{0}$. \ These ellipses are unbounded in
number, and their three-frames can rotate independently, subject only to the
restrictions of continuity. \ A practical, albeit not fundamental, further
restriction is introduced by the fact that sufficiently close to the C point a
linear expansion of the field suffices to determine what happens in the
generic case. \ As becomes apparent, this leads to additional interconnections
between indices, and forbidden combinations of indices (selection rules), so
that structures that are allowed in principle by geometrical and topological
constraints may not appear in practice.%

\subsection{Computed Optical Fields}

\subsubsection{Random Speckle Field}

We study here two types of computer simulated random optical fields. \ The
first is a field composed of a large number of randomly phased, linearly
polarized plane waves with random polarization and random propagation
directions. \ This speckle field (speckle pattern), described in full detail
in [$13-15$], has the advantage that it is an is an exact solution of
Maxwell's equations. \ It has two disadvantages, however: The first is that C
lines must be found and followed empirically, using a time consuming search
based on either the zeros of $\mathbf{E}\boldsymbol{\cdot}\mathbf{E}$, or the
discriminant $D_{C}$ of the coherency matrix described above. \ Thus, the
number of different lines that can be studied is limited, and as a result, a
limited menu of different M\"{o}bius strips is available for detailed study.
\ Nonetheless, this random field serves as our gold standard, and is used to
verify the existence of the structures measured with the aid of the linear
expansion described below.

\subsubsection{Linear Expansion}

Locating the origin at an arbitrary point in a generic complex optical field,
the three Cartesian components of the field in the immediate vicinity of the
point can be expanded to first order (the order required here) as%
\begin{equation}
\left(
\begin{array}
[c]{c}%
E_{x}\\
E_{y}\\
E_{z}%
\end{array}
\right)  =\left(
\begin{array}
[c]{c}%
a_{x}+\text{i}b_{x}\\
a_{y}+\text{i}b_{y}\\
a_{z}+\text{i}b_{z}%
\end{array}
\right)  +\left(
\begin{array}
[c]{ccc}%
P_{xx}+\text{i}Q_{xx} & P_{xy}+\text{i}Q_{xy} & P_{xz}+\text{i}Q_{xz}\\
P_{yx}+\text{i}Q_{yx} & P_{yy}+\text{i}Q_{yy} & P_{yz}+\text{i}Q_{yz}\\
P_{zx}+\text{i}Q_{zx} & P_{zy}+\text{i}Q_{zy} & P_{zz}+\text{i}Q_{zz}%
\end{array}
\right)  \left(
\begin{array}
[c]{c}%
x\\
y\\
z
\end{array}
\right)  , \label{GenFieldXpan}%
\end{equation}
where, for example, $P_{xy}=\operatorname{Re}\left(  \partial E_{x}/\partial
y\right)  _{0}$, $Q_{xy}=\operatorname{Im}\left(  \partial E_{x}/\partial
y\right)  _{0}$, etc. \ The divergence condition $\mathbf{\nabla
}\boldsymbol{\cdot}\mathbf{E}=0$ can be satisfied by setting $P_{zz}=-\left(
P_{xx}+P_{yy}\right)  $, and $Q_{zz}=-\left(  Q_{xx}+Q_{yy}\right)  $.

At the selected point, which can have arbitrary elliptical polarization, we
erect a convenient coordinate system in which the $+z$-axis is along the
positive normal $\boldsymbol{\gamma}$ to the plane of the ellipse, the
$x$-axis is along the major ellipse axis $\boldsymbol{\alpha}$, and the
$y$-axis is along the minor axis $\boldsymbol{\beta}$. \ In this coordinate
system the plane $z=0$ is the principal plane $\Sigma_{0}$ of the ellipse at
the origin.

When the point at the origin is a C point on a C line, the condition $\left(
\mathbf{E}\boldsymbol{\cdot}\mathbf{E}\right)  _{0}=0$ can be satisfied by
setting $a_{y}=a_{z}=0$, $b_{x}=b_{z}=0$, and $a_{y}=\pm\left\vert
a_{x}\right\vert $. \ Thus, in $\Sigma_{0}$ all constraints are satisfied
without loss of generality by writing for a C point
\begin{subequations}
\label{CpointE}%
\begin{align}
E_{x}  &  =a+\left(  P_{xx}+\text{i}Q_{xx}\right)  x+\left(  P_{xy}%
+\text{i}Q_{xy}\right)  y,\label{CpointE_a}\\
E_{y}  &  =\text{i}s\left\vert a\right\vert +\left(  P_{yx}+\text{i}%
Q_{yx}\right)  x+\left(  P_{yy}+\text{i}Q_{yy}\right)  y,\label{CpointE_b}\\
E_{z}  &  =\left(  P_{zx}+\text{i}Q_{zx}\right)  x+\left(  P_{zy}%
+\text{i}Q_{zy}\right)  y, \label{CpointE_c}%
\end{align}
where $s=\pm1$, and $a$ and the $P$ and $Q$ remaining in Eq. (\ref{CpointE})
are unconstrained.

Simpler expansions are also possible. \ The one-half-twist Mobius strip in
Fig. \ref{Fig1}a \ is closely approximated by $E_{x}=-1-\left(  1-i\right)
y,E_{y}=-i+x,E_{z}=-ix+y$, the three-half-twist strip in Fig. \ref{Fig1}b by
$E_{x}=1-y,E_{y}=i-x,E_{z}=y$.

We make contact with the random speckle field described above by generating
the various constants in Eq. (\ref{CpointE}) with the aid of the procedure
described below: this procedure is appropriate to a random field whose
components obey circular Gaussian statistics, and ensures that the M\"{o}bius
strips we study have generic properties that are likely to correspond to those
in real physical fields.

We start by writing the joint probability density function (PDF)
$\mathfrak{P}$ of the three Cartesian field components in Eq.
(\ref{GenFieldXpan}), which describes a general point in the fixed laboratory
$xyz$ frame,\ as
\end{subequations}
\begin{subequations}
\label{Gauss}%
\begin{align}
\mathfrak{P}\left(  E_{x},E_{y},E_{z}\right)   &  =\mathfrak{P}\left(
E_{x}\right)  \mathfrak{P}\left(  E_{y}\right)  \mathfrak{P}\left(
E_{z}\right)  ,\label{Gauss_a}\\
\mathfrak{P}\left(  E_{j}\right)   &  =\mathfrak{P}\left(  a_{j},b_{j}%
,P_{jx},...,Q_{jz}\right)  ,\;j=x,y,z\label{Gauss_b}\\
\mathfrak{P}\left(  a_{j},b_{j},P_{jx},...,Q_{jz}\right)   &  =\mathfrak{P}%
_{G}\left(  a_{j}\right)  \mathfrak{P}_{G}\left(  b_{j}\right)  \mathfrak{P}%
_{G}\left(  P_{jx}\right)  ...\mathfrak{P}_{G}\left(  Q_{jz}\right)
,\label{Gauss_c}\\
\mathfrak{P}_{G}\left(  u\right)   &  =\frac{1}{\sqrt{2\pi\left\langle
u^{2}\right\rangle }}\exp\left(  -\frac{u^{2}}{2\left\langle u^{2}%
\right\rangle }\right)  . \label{Gauss_d}%
\end{align}

But we want $\mathfrak{P}\left(  E_{x},E_{y},E_{z}\right)  $ not at a general
point, but at a special point, a C point, and not in the fixed laboratory
$xyz$ frame, but in the principal axis $x^{\prime}y^{\prime}z^{\prime}$ frame
of the C point. \ Accordingly, we proceed numerically as follows. \ In accord
with Eq. (\ref{Gauss}) we first choose the sign of $s$ and the values of the
various $a,b,P,$ and $Q$ in Eq. (\ref{GenFieldXpan}) by consulting a random
number generator that produces a Gaussian distribution with, for convenience,
unit variance. \ We then adjust the parameters $a$ and $b$ so as to satisfy
$\left(  \mathbf{E}\boldsymbol{\cdot}\mathbf{E}\right)  _{0}=0$, writing
$a_{z}$ and $b_{z}$ in terms of $a_{x},a_{y},b_{x},b_{y}$. Next, we find the
eigenvectors of the coherency matrix in Eq. (\ref{RealCohMatrix}); these
eigenvectors yield the direction cosines of the principal axes $x^{\prime
}y^{\prime}z^{\prime}$ of the C point relative to the laboratory $xyz$ frame.
\ As discussed above, eigenvector $\boldsymbol{\nu}_{1}$ is the well defined
normal to the C circle (axis $\boldsymbol{\gamma}$), whereas $\boldsymbol{\nu
}_{2}$ and $\boldsymbol{\nu}_{3}$ are arbitrary orthogonal directions in the
plane of the C circle. \

We form a matrix $\mathbf{D}$ from these direction cosines, and use it to
transform the parameters $a$, $b$, $P$, and $Q$, in the laboratory frame to a
parameter set $a^{\prime}$, $b^{\prime}$, $P^{\prime}$, and $Q^{\prime}$ in
the principal axis frame of the C point. \ Forming a vector $\mathbf{a}$
(vector $\mathbf{b})$ from the $a$ ($b$) parameters, and matrices $\mathbf{P}$
and $\mathbf{Q}$ from the $P$ and $Q$ parameters, respectively, and
transforming both coordinates and field components, we obtain
\end{subequations}
\begin{subequations}
\label{D matrix}%
\begin{align}
\mathbf{a}^{\prime}  &  =\mathbf{D\cdot a,}\label{D_a}\\
\mathbf{b}^{\prime}  &  =\mathbf{D\cdot b,}\label{D_b}\\
\mathbf{P}^{\prime}  &  =\mathbf{D\cdot P\cdot D}^{-1},\label{D_c}\\
\mathbf{Q}^{\prime}  &  =\mathbf{D\cdot Q\cdot D}^{-1}. \label{D_d}%
\end{align}

As expected, we find $a_{z}^{\prime}=b_{z}^{\prime}=0$. \ Writing
$a_{x}^{\prime}+\,$i$b_{x}^{\prime}=A_{x}^{\prime}\exp\left(  \text{i}%
\varphi_{x}^{\prime}\right)  $, $a_{y}^{\prime}+\,$i$b_{y}^{\prime}%
=A_{y}^{\prime}\exp\left(  \text{i}\varphi_{y}^{\prime}\right)  $, we find,
again as expected, $A_{x}^{\prime}=A_{y}^{\prime}$ and $\left\vert \varphi
_{x}^{\prime}-\varphi_{y}^{\prime}\right\vert =\pi/2$. \ But this is not yet
in the convenient form of Eq. (\ref{CpointE}). \ We complete the
transformation by writing $\mathbf{E}^{\prime\prime}=\exp(-ib_{x}^{\prime
}/a_{x}^{\prime})\mathbf{E}^{\prime}$, extract the real and imaginary parts of
$\mathbf{E}^{\prime\prime}$, set $z=0$, and after dropping for convenience the
double prime superscripts obtain Eq. (\ref{CpointE})

The individual PDFs of $P^{\prime}$ and $Q^{\prime}$ remain circular Gaussians
with unit variance. \ This is not surprising, because $P$ and $Q$ are
transformed under a similarity transformation in which the arguments of the
cosines in $\mathbf{D}$ are uniformly distributed $0-2\pi$. \ The above
procedure is required because it preserves specific correlations between the
final $P^{\prime\prime}$ and $Q^{\prime\prime}$ that are absent in the initial
$P$ and $Q$; these correlations arise because the elements of $\mathbf{D}$ are
themselves functions of $P$ and $Q$ [$15$].

In contrast to the PDFs of $P$ and $Q$, the PDF of $a$, shown in Fig. 4,
differs markedly from a Gaussian.%

\begin{figure}
[h]
\includegraphics[width=0.5\textwidth]%
{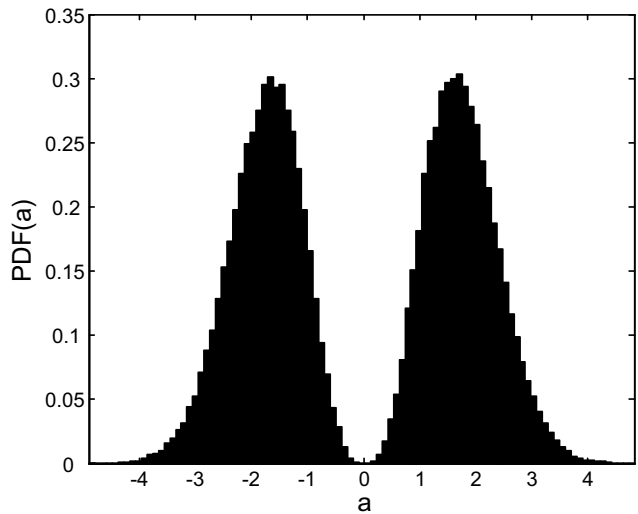}%
\caption{Final PDF of $a$ in Eq. (\ref{CpointE}).}%
\label{Fig4}%
\end{figure}

Although Eq. (\ref{CpointE}) appears to describe an isolated C point at the
origin, this C point is, in fact, part of a continuous C line. \ In addition
to the C point at the origin, Eq. (\ref{CpointE}) can also produce one or more
parasitic C points (C lines) at locations that depend upon the various
parameters. \ As long as the surrounding circle $\sigma_{0}$ is sufficiently
small that it does not contain, or come close to, these parasitic C points
they cause no trouble. \ Occasionally, with a probability of $\sim1-2\%$, a
parasitic C point is so close to the C point at the origin that it strongly
distorts the field in the region of the central point. \ When this happens the
M\"{o}bius strip at the origin can become so distorted that its parameters
cannot be accurately measured. \ In Section IV where we discuss the statistics
of the strips we eliminate such cases (outliers) from the analysis.

\subsection{Graphics}

We discuss here three important aspects of the graphic representation of the
M\"{o}bius strips and other wavefield structures that are presented throughout
this report.

\subsubsection{Length scales}

We start by noting that there is no geometric or, indeed any, relationship
between the radius of $\sigma_{0}$, which is a true length, and the length of
the line used to represent an ellipse axis.

The length scale of $\sigma_{0}$ is set by the variances of the Gaussian
distributions in Eq. (\ref{Gauss}) that are used for the constants $a$ and
$b$, which measure the field amplitude, and the derivatives $P$ and $Q$, that
measure the change in field strength per unit length. \ We take the field and
derivative variances to be unity, which fixes the length scale of $\sigma_{0}$
to also be unity. \ When the radius $r_{0}$ of $\sigma_{0}$ is sufficiently
small (i.e. small compared to one), the field scales uniformly with $r_{0}$
and its structure becomes independent of this radius. \ Here, we take
$r_{0}=10^{-4}$, having verified that this is \textquotedblleft sufficiently
small\textquotedblright\ in the above sense.

Although the length scale of the field is also unity, the length scale of the
polarization ellipse remains arbitrary. \ The reason is that the ellipse is a
representation of the magnitude and direction of the electric field vector
over an optical cycle \emph{at a point}. \ We can therefore choose any
convenient scale to represent the ellipse, because in the physical field two
arbitrarily close ellipses do not overlap. \ Thus, the widths of the ribbons
representing the M\"{o}bius strips in Figs. \ref{Fig1} and \ref{Fig2}, and in
all other figures presented here, are adjusted as needed for clarity. \ Of
course, in a given figure a single length scale is used for all ellipses.

\subsubsection{Scaling}

By continuity, the tilt of the plane of the ellipses on $\sigma_{0}$ relative
to the plane $\Sigma_{0}$, the plane of the C circle, decreases to zero as the
radius $r_{0}$ goes to zero. \ This tilt is of order $r_{0}$, which for
$r_{0}=10^{-4}$ is of order $0.005^{o}$. \ In later figures (such as Figs.
\ref{Fig7} $-$ \ref{Fig12}) we will show the projections of the ellipse axes
onto the $XZ$-plane or the $YZ$-plane of Fig. \ref{Fig3}, i.e. the planes
$\tau_{0}$ and $\pi_{0}$. \ In such figures we use anisotropic scaling,
expanding the $Z$-axis relative to the $X$- or $Y$-axis so as to make visible
the \emph{relative} tilts of the axes of different ellipses, the absolute
values of these tilts being unimportant in these figures.

\subsubsection{Paths}

As mentioned, we choose $\sigma_{0}$ to be a circle centered on the C point
for simplicity $-$ other simple paths such as ellipses centered on the C point
yield the same values for the topological indices as does $\sigma_{0}$. \ The
path must be \textquotedblleft centered \textquotedblright\ because when
shrunk to zero it must enclose the C point

The choice of a circle centered on the C point yields the most symmetrical
possible form for the M\"{o}bius strips, enhancing our ability to understand
their sometimes complex structures. \ But even within the domain of circles
there is nothing special about a radius of say $r_{0}=10^{-4}$ $-$
$r_{0}=1.1\times10^{-4}$, or any other value, is obviously equally good.
\ From this follows that the C point is surrounded by an infinite set of
centered nested M\"{o}bius strips.

It is, of course, difficult, if not impossible, to visualize the full 3D
structure of such a field. \ Below we dissect out a single representative
M\"{o}bius strip of the nested strips that surrounds C points, the $\sigma
_{0}$ M\"{o}bius strip, and proceed to study this strip in detail.

\section{M\"{O}BIUS STRIP INDICES}

Throughout this report we restrict ourselves to the case in which the plane of
observation $\Sigma$ coincides with the principal plane $\Sigma_{0}$. \ In
this plane we find that all indices are the same for axes $\boldsymbol{\alpha
}$ and $\boldsymbol{\beta}$, even though the detailed geometries of the
M\"{o}bius strips generated by these axes are different. For simplicity and
uniformity, in what follows all examples presented are for axis
$\boldsymbol{\alpha}$. \ As $\Sigma$ rotates away from $\Sigma_{0}$ a complex
set of phenomena set in: the universal equivalence between axes
$\boldsymbol{\alpha}$ and $\boldsymbol{\beta}$ is broken, right-handed
M\"{o}bius strips transform into left-handed one, and vice versa,
one-half-twist strips change to three-half twist strips, and vice versa,
indices abruptly change sign, etc. \ These phenomena will be reported on separately.

\subsection{Topological Indices of the Projection onto ${\protect\Large \Sigma
}_{0}$}

\subsubsection{Indices $I_{\alpha,\beta}$}

As indicated in the Introduction, the sole topological index used in previous
studies to characterize C points and C lines is based on the rotation about
the C point of the projections onto $\Sigma$ of the major or minor axes of the
ellipses on the surrounding circle $\sigma$. \ When $\Sigma=\Sigma_{0}$ (so
that $\sigma=\sigma_{0}$) it is easily seen that index $I$ (Fig. \ref{Fig2})
is the same for both axis \textbf{$\boldsymbol{\alpha}$} and axis
\textbf{$\boldsymbol{\beta}$}: \ sufficiently close to the C point the
surrounding ellipses are tilted negligibly out of the plane $\Sigma_{0}$, the
plane of the C circle, so that the projections of $\boldsymbol{\alpha}$ and
$\boldsymbol{\beta}$ onto $\Sigma_{0}$ are very nearly orthogonal at all
points on $\sigma_{0}$. \ Thus, these projections are locked together, and as
one rotates so does the other: as a result, $I_{a}=I_{\boldsymbol{\beta}}$.
\ This is true even when $\Sigma\neq\Sigma_{0}$. \ As expected, we find that
in our simulations $I_{a}=I_{\boldsymbol{\beta}}$ in all cases ($\Sigma
=\Sigma_{0}$, and $\Sigma\neq\Sigma_{0}$): accordingly in what follows we
write this index as $I_{a,\boldsymbol{\beta}}$.

Generically, $I_{a,\boldsymbol{\beta}}=\pm1/2$ [$1,3$]. \ This is not due to
geometric or topological constraints, but is rather a consequence of the linear
expansion; when higher order terms dominate, higher order values for this
index are possible [$32$]. \ Half integer values for $I_{a,\boldsymbol{\beta}%
}$ can occur because of the symmetry of the ellipse which returns to itself
after rotation by $\pi$; for the rotations of a vector, for example, only
integer values for the index are possible.

Hidden in the diagrams that illustrate $I_{a,\boldsymbol{\beta}}$ lie long
overlooked clues to the fact that these 2D diagrams are projections of 3D
M\"{o}bius strips. \ These clues are discussed in Fig. \ref{Fig5}\ %

\begin{figure}
[h]
\includegraphics[width=0.7\textwidth]%
{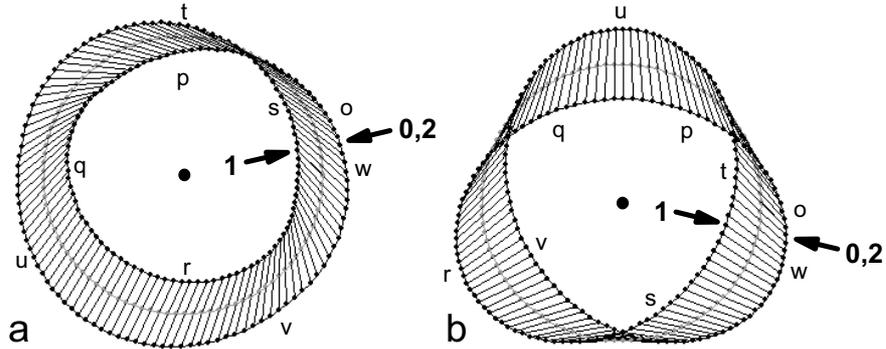}%
\caption{Hidden clues. \ As shown in Fig. \ref{Fig2}, in determining
$I_{\alpha,\beta}$ one traditionally focuses on the rotation of the
projections onto $\Sigma_{0}$ of the major or minor axes of the ellipses on
the surrounding circle $\sigma_{0}$. \ It is amusing to note in retrospect
that these diagrams contain an important clue overlooked during the twenty
five years since their inception. \ If instead of focusing on the rotation of
the axes (here, as in Fig. \ref{Fig2}, the thin straight lines) one focuses
instead on the axis endpoints (small black dots), it immediately becomes
apparent that the flat ribbon formed by the axes has, just like the parent
M\"{o}bius strip, only one edge. \ Starting at the point labeled \textbf{0}
and proceeding counterclockwise along the edge (o $\rightarrow$ p
$\rightarrow$ ... $\rightarrow$ v $\rightarrow$ w), after one $360^{o}$
circuit around the C point (central large black dot) one reaches the point
labeled \textbf{1}. \ Continuing along a second circuit one returns to the
point labeled \textbf{2}, which, because it is coincident with the starting
point \textbf{0}, demonstrates that the axis ribbon has only a single edge.
\ The single self intersection of the curve formed by the endpoints in (a),
which is Fig. \ref{Fig2}a redrawn, suggests (but of course does not prove) the
possible existence of a one-half-twist M\"{o}bius strip, whereas the three
self intersections of this curve in (b), which is Fig. \ref{Fig2}d redrawn,
suggests (but, again, does not prove - see Fig. \ref{Fig2}b) the possible
existence of a three-half-twist M\"{o}bius strip. \ }%
\label{Fig5}%
\end{figure}

\subsubsection{Index $I_{\gamma}$}

What about the projection onto $\Sigma_{0}$ of axis $\boldsymbol{\gamma}$ of
the ellipses on $\sigma_{0}$? \ As shown in Fig. \ref{Fig6}, these projections
rotate about the C point with integer winding number $I_{\gamma}=$ $\pm1$.
\ $I_{\gamma}=\pm1$ because $\boldsymbol{\gamma}$ is a vector. \ But wait!
\ Axis $\boldsymbol{\gamma}$ is well defined at the C point, and normally we
associate winding numbers with \emph{singularities}, i.e. with properties that
are \emph{undefined}; for example winding number $I_{a,\boldsymbol{\beta}}$ is
associated with the fact that axes $\boldsymbol{\alpha}$ and
$\boldsymbol{\beta}$ are undefined at the C point. \ So is $I_{\gamma}$ a
valid topological index, or not?

In answering this question in the affirmative we note the following: Although
axis $\boldsymbol{\gamma}$ is well defined, its projection onto $\Sigma_{0}$
is singular. \ The reason is that although the projection of this axis for all
the surrounding ellipses are lines with well defined directions, the
projection of axis $\boldsymbol{\gamma}$ for the C point itself is a point
whose direction is undefined. \ Thus, in $\Sigma_{0}$, axis
$\boldsymbol{\gamma}$ and axes $\boldsymbol{\alpha}$ and $\boldsymbol{\beta}$
behave similarly, both projections have undefined directions $-$ axes
$\boldsymbol{\alpha}$ and $\boldsymbol{\beta}$ because the C circle has no
preferred direction, axis $\boldsymbol{\gamma}$ because a point (a circle of
vanishingly small radius) also has no preferred direction.

The above comparison is closer still if instead of plotting the projections of
the ellipses we plot the quantities given in Eq. (\ref{alphabetagamma}). \ As
defined there, $\boldsymbol{\alpha}$ and $\boldsymbol{\beta}$ go to zero at a
C point, and maps of the projections of $\boldsymbol{\alpha}$,
$\boldsymbol{\beta}$, and $\boldsymbol{\gamma}$, are similar $-$ all three
maps show a point at the location of the C point surrounded by lines that
rotate about the point with winding number of $\pm1/2$ ($\pm1$) for axes
$\boldsymbol{\alpha}$ and $\boldsymbol{\beta}$ (axis $\boldsymbol{\gamma}$).

Finally, we recall the well known maxim that the index of a path (here the
circle $\sigma_{0}$) is a property of the path. \ The index of the path
becomes the index of the C point by a limiting process in which we shrink the
path onto the point and observe that the index remains invariant under this
transformation. \ $I_{\gamma}$ satisfies this criterion also.%

\begin{figure}
[h]
\includegraphics[width=0.75\textwidth]%
{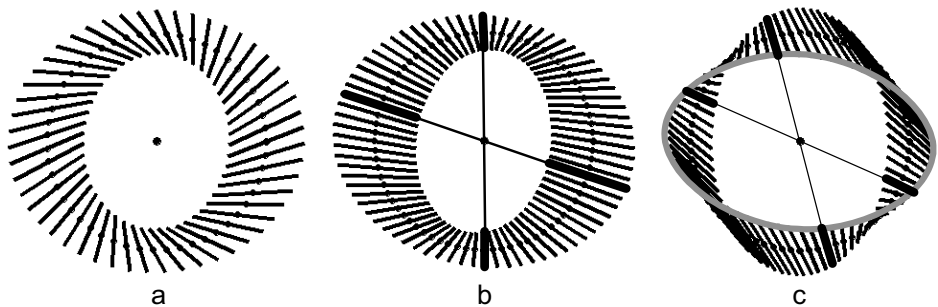}%
\caption{Index $I_{\gamma}$. \ (a) and (b) $I_{\gamma}=+1$. \ (c) $I_{\gamma
}=-1$. \ Here the gray ellipse facilitates following the retrograde rotation
of the axes. \ The thick black markers and radial lines in (b) and in (c)
relate to the line classification, and are discussed in Section III.D.2.}%
\label{Fig6}%
\end{figure}

What is the three dimensional structure generated by axis $\boldsymbol{\gamma
}$? \ As can be seen in Fig. \ref{Fig7} it is a segment of a cone, not a
M\"{o}bius strip. \ This cone is analogous to the cones described previously
that surround ordinary, i.e. nonsingular, points [$14,15$].%

\begin{figure}
[h]
\includegraphics[width=0.75\textwidth]%
{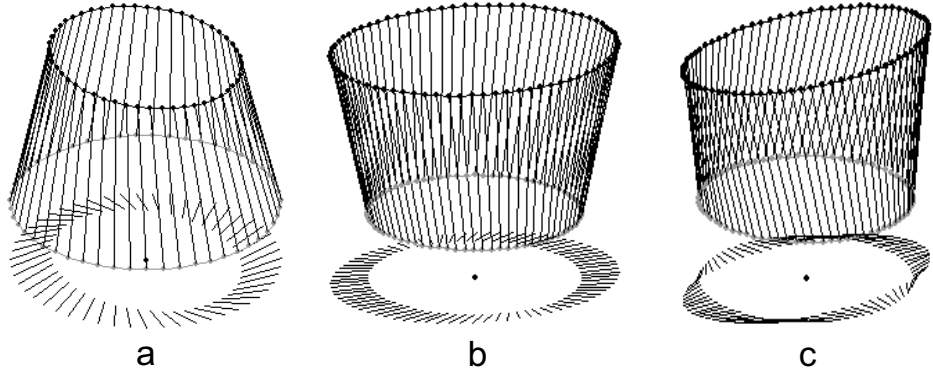}%
\caption{$\boldsymbol{\gamma}$ cones. \ (a), (b), and (c), respectively, are
3D views of the cones that produce projections (a), (b), and (c) in Fig.
\ref{Fig6}.}%
\label{Fig7}%
\end{figure}

\subsection{Topological Indices of the Projection onto ${\protect\Large \tau
}_{0}$}

The classic construction of a M\"{o}bius strip, known to middle-school
children everywhere, is as follows: take a strip of paper, twist one end
through a half turn, and glue the two ends together. \ A similar construction
serves to model the optical M\"{o}bius strips discussed here: attach short
rods (representing ellipse axes $\boldsymbol{\alpha}$ or $\boldsymbol{\beta}$)
at their midpoints to a strip of flexible material, introduce one or three
half-twists, and connect together the ends of the twisted strip.

The strip can be twisted in one of two senses to form either a right- or
left-handed screw. \ Rotating the ends of the strip to bring them together can
also be done in one of two ways, when viewed from, say, above, either
clockwise or counterclockwise. But the geometry and topology of the strip
remain the same whichever choice (clockwise or counterclockwise) is made, and
only the sense (right- or left-handed) and number (one or three) of the
initial half-twists determines the nature of the strip.

In the remainder of this report we drop for convenience the subscript
\textquotedblleft$0\textquotedblright$ on $r_{0\text{,}}$ $X_{0}$, $Y_{0}$,
$Z_{0}$, and $\chi_{0}$, its presence being understood.

\subsubsection{Indices $\tau_{\alpha,\beta}$}

Following the (unfortunate) convention mandated for circularly polarized
light, namely that the left-handed (right-handed) screw traced out in space by
the rotation of the electric vector as the light propagates is labelled
right-handed (left-handed) polarization, we attach a positive (negative) sign
to left (right) handed M\"{o}bius strips. \ Adding in the number of half
twists, we find four possible values for the twist index $\tau$, $\tau
=\pm1/2,\pm3/2$.

$\tau$ can be obtained by following the rotation of the ellipse axes projected
onto the rotating plane $\tau_{0}$. \ We label this projection $\mathbf{T}$,
and illustrate the procedure in Fig. \ref{Fig8}. \ Here the angle $\zeta$
relative to the $X$-axis is measured as $\zeta\left(  \chi\right)
=\arctan\left(  T_{Z}\left(  \chi\right)  ,T_{X}\left(  \chi\right)  \right)
$, where, Fig. \ref{Fig3}a, $\chi$ measures position on $\sigma_{0}$.
\ $\zeta\left(  \chi\right)  $ is unfolded (unwrapped) as required, and $\tau$
is computed as $\tau=\Delta\zeta/2\pi$, where $\Delta\zeta=\zeta\left(
2\pi\right)  -\zeta\left(  0\right)  $.%

\begin{figure}
[h]
\includegraphics[width=0.75\textwidth]%
{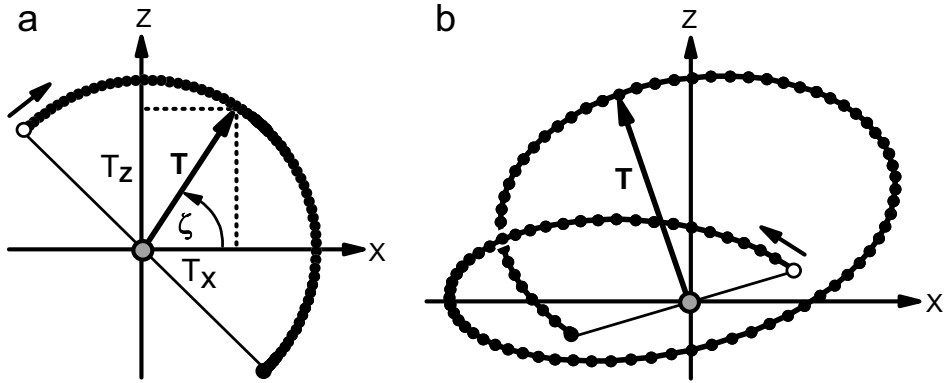}%
\caption{Index $\tau_{\alpha,\beta}$. \ (a) M\"{o}bius strip with a single
half-twist. \ This strip is the one shown in Fig. \ref{Fig1}a. \ The origin of
vector $\mathbf{T}$ is on $\sigma_{0}$ (gray circle) whereas its head
terminates on the projections onto $\tau_{0}$ of the endpoints of the
surrounding ellipse axes (small black dots). $\ \mathbf{T}$ begins a $2\pi$
circuit on $\sigma_{0}$ at the the small white circle, and in the case shown
here rotates clockwise through $180^{o}$ to the large black dot that
terminates the curve of endpoints. Here and throughout, this black dot is
connected by a thin straight line to the initial white dot in order to
emphasize that these two points terminate opposite ends of the \emph{same}
ellipse axis. \ The position of $\mathbf{T}$ relative to the $X$-axis is
measured by angle $\zeta=\arctan\left(  T_{Z},T_{X}\right)  $. \ At the start
(end) of the circuit shown here $\zeta=135^{o}$ ($\zeta=-45^{o}$). \ The net
rotation is therefore $\Delta\zeta=-45^{o}-135^{o}=-180^{o}$, and
$\tau_{\alpha,\beta}=-1/2$. \ (b) M\"{o}bius strip with three half-twists.
\ This strip is the one shown in Fig. \ref{Fig1}b. \ Here $\mathbf{T}$ rotates
counterclockwise through $+540^{o}$ over a $2\pi$ circuit on $\sigma_{0}$,
rotating by a little less than $360^{o}$ over the lower loop and by a little
more than $180^{o}$ over the upper one. \ $\tau_{\alpha,\beta}$ is therefore
$\tau_{\alpha,\beta}=+3/2$. \ The starting point on $\sigma_{0}$ is arbitrary,
and choosing a different starting point for the circuit changes the initial
and final values of $\zeta$, but not their difference $\Delta\zeta$.
\ Although by convention we traverse $\sigma_{0}$ in the counterclockwise
direction as viewed from the positive end of $\boldsymbol{\gamma}$, the $+z$
direction in Fig. \ref{Fig3}a, traversing $\sigma_{0}$ in the opposite,
clockwise, direction does not change either the sign or magnitude of
$\tau_{\alpha,\beta}$. This reflects the fact that the handedness of a screw
$-$ right or left $-$ remains the same when the screw is viewed from any
direction.}%
\label{Fig8}%
\end{figure}

Although not an obvious geometrical or topological necessity, we find in
$\Sigma_{0}$ (and only in $\Sigma_{0}$) that in all cases the values of $\tau$
obtained for axes \textbf{$\boldsymbol{\alpha}$} and
\textbf{$\boldsymbol{\beta}$} are always the same, even though the detailed
geometry of the \textbf{$\boldsymbol{\alpha}$} and \textbf{$\boldsymbol{\beta
}$} Mobius strips may differ substantially. \ Accordingly, here we label this
index $\tau_{\alpha,\beta}$. \ In Section III.D.1 we show that within the
linear approximation only one- and three-half-twist M\"{o}bius strips are
possible, so that generically $\tau_{\alpha,\beta}=\pm1/2,\pm3/2$.

\paragraph{Phase ratchet rules}

As $\mathbf{T}$ rotates its endpoint traces out one of the the curves shown in
Fig. \ref{Fig8}. \ The signed (plus for counterclockwise, minus for clockwise)
crossings of this curve with the $X$ and $Z$-axes can be used to determine
$\tau_{\alpha,\beta}$ by means of a minor variation of what we called the
\textquotedblleft phase ratchet rules\textquotedblright\ [$31$]. \ For a
one-half-twist M\"{o}bius strip in a $2\pi$ ($4\pi$) circuit on $\sigma_{0}$,
Fig. \ref{Fig8}a, the vector $\mathbf{T}$ rotates through $\pi$ ($2\pi$), and
the endpoint curve crossing sequence is $XZ$ ($XZXZ$), or its cyclic
permutation. \ For a three-half-twist strip $\mathbf{T}$ rotates through
$3\pi$ ($6\pi$), Fig. \ref{Fig8}b, and the endpoint curve crossing sequence is
$XZXZXZ$ ($XZXZXZXZXZXZ$), or its cyclic permutation. \ An important
complication, however is that the rotation of $\mathbf{T}$ is generally not
monotonic, and $\mathbf{T}$ can, and often does, wander back and forth during
its overall rotation, leading to additional crossings of the endpoint curve
with the $XZ$-axes. \ Here the phase ratchet rules come into play: these rules
state that adjacent terms of the same axis in the crossing sequence (these
terms necessarily have opposite signs) are to be erased; thus, the sequence
$XZZZ$ for a $2\pi$ circuit, for example, is reduced to $XZ$, indicating that
the M\"{o}bius strip has a single half-twist. \ These rules are illustrated in
Fig. \ref{Fig9}.%

\begin{figure}
[h]
\includegraphics[width=0.75\textwidth]%
{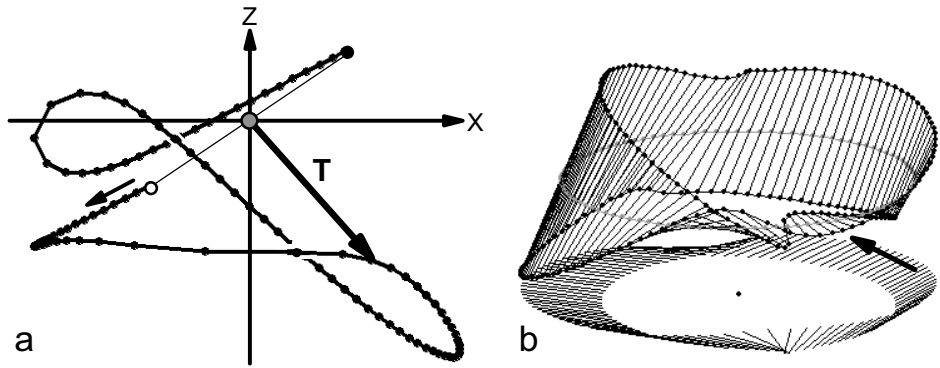}%
\caption{Phase ratchet rules. \ (a) Endpoint curve. \ (b) M\"{o}bius strip.
\ For the M\"{o}bius strip shown in (b) the endpoint curve in (a) is a
complicated figure with multiple crossings of the $X$ and $Z$ axes. \ As a
result, the rotation of the vector $\mathbf{T}$ is so complicated that its net
angle of rotation cannot be determined visually. \ The phase ratchet rules
simplify the calculation and permit $\tau_{\alpha,\beta}$ to be obtained by
inspection. \ In applying the rules one first lists the signed crossings of
the endpoint curve with the $X$ and $Z$ axes in the order that they occur as
one traverses the curve. \ Starting at the small white circle and proceeding
along the curve, the sequence of signed crossings is $Z_{+}Z_{-}X_{-}%
X_{+}X_{-}Z_{-}$, where subscript $+$ ($-$) indicates that $\mathbf{T}%
$\textbf{ }rotates counterclockwise (clockwise) at the crossing. \ Applying
the phase ratchet rules one erases adjacent terms in the same axis, in this
case the terms $Z_{+}Z_{-}$\ and $X_{-}X_{+}$,\ or $X_{+}X_{-}$, to obtain
$X_{-}Z_{-}$. \ From this canonical sequence it immediately follows, Fig.
\ref{Fig8}a, that $\tau_{\alpha,\beta}=-1/2$. \ The Mobius strip shown in (b)
can be generated from an ordinary right-handed half-twist strip by grasping a
segment of the ribbon between thumb and forefinger and twisting the segment by
$\sim180^{o}$ (see the arrow in (b)). \ This introduces a pair of twists $-$
one right-handed and the other left-handed. \ Although this operation changes
the geometry of the strip, because the new right and left handed twists cancel
the topology remains that of a right-handed one-half-twist M\"{o}bius strip.
\ As discussed in Section IV, in a random optical field a\ surprisingly large
fraction of all M\"{o}bius strips have complicated structures similar to the
one shown here. \ \ }%
\label{Fig9}%
\end{figure}

\subsubsection{Indices $d\tau_{\alpha,\beta}$}

As is evident from the phase ratchet rules, the complicated shape of the
endpoint curve in Fig. \ref{Fig9}a is not fully characterized by $\tau
_{\alpha,\beta}$; here we discuss a second topological invariant that adds
additional information about such curves. \ (A complete characterization of
the curve would require, of course, an infinite set of indices or moments.)

A standard, widely used characterization of curves in the plane is the
Poincar\'{e} index. \ This index measures the winding number of the
\emph{tangent} to the curve, in our case the endpoint curve measured over one
$2\pi$ circuit of $\sigma_{0}$. \ Here we denote an analog of this index by
$d\tau$, and calculate it as $d\tau=\arctan\left(  dT_{Z}/ds,dT_{X}/ds\right)
$ using the finite difference approximation $d\tau\left(  \chi_{j}\right)
=\arctan\left(  T_{Z}\left(  \chi_{j+1}\right)  -T_{Z}\left(  \chi_{j}\right)
,T_{X}\left(  \chi_{j+1}\right)  -T_{X}\left(  \chi_{j}\right)  \right)  $
with uniform increment $\Delta s_{j}=s_{j+1}-s_{j}=$ $\Delta s>0$, where, Fig.
\ref{Fig3}a, $s$ measures position on $\sigma_{0}$.

Although the endpoint curves differ for axes $\boldsymbol{\alpha}$ and
$\boldsymbol{\beta}$, and although, like for $\tau_{\alpha,\beta}$, there is
no obvious geometrical or topological necessity for $d\tau$ to be the same for
both axes, we find, just like for $\tau_{\alpha,\beta}$, that in $\Sigma_{0}$
(and again, only in $\Sigma_{0}$) in all cases $d\tau_{\boldsymbol{\alpha}%
}=d\tau_{\boldsymbol{\beta}}$. \ Accordingly, in what follows we label this
index $d\tau_{\alpha,\beta}$.

Like $\tau_{\alpha,\beta}$, $d\tau_{\alpha,\beta}$ takes on the values
$d\tau_{\alpha,\beta}=\pm1/2,\pm3/2$. \ We show in Section IV that only ten of
the sixteen possible combinations of $\tau_{\alpha,\beta}$ and $d\tau
_{\alpha,\beta}$ appear, so that these two indices are not completely
independent one of the other. \ An example in which $\tau_{\alpha,\beta}=-1/2$
and $d\tau_{\alpha,\beta}=+3/2$ is shown in Fig. \ref{Fig10}. \ As shown in
this figure, we take the sign of the local rotation of the tangent vector to
be $+$ ($-$) if the vector rotates locally in the counterclockwise (clockwise) direction.

But if $d\tau_{\alpha,\beta}$ is even partially independent of $\tau_{\alpha,\beta}$, how can
we be sure that $d\tau_{\alpha,\beta}$ takes on only the half integer values
$\pm1/2,\pm3/2$? \ The answer consists of two parts: in the first part, given
below, we show that over a single $2\pi$\ circuit of $\sigma_{0}$ the vector
tangent to the endpoint curve must rotate through $n\pi$, where $n$ is an odd
positive/negative integer. \ In the second part, which still remains to be
accomplished, one shows that within the linear approximation $\left\vert
n\right\vert =1,3$.

Over a $4\pi$ circuit of $\sigma_{0}$, when we return to our starting point
the tangent vector to the endpoint curve must also return to itself, so for a
$4\pi$ circuit the tangent vector rotates through $2n\pi$. \ But as
illustrated in Fig. \ref{Fig10}, the endpoint curve for the $4\pi$ circuit has
inversion symmetry, so that over any $2\pi$ circuit the vector rotates through
$n\pi$. \ The inversion symmetry of the endpoint curve reflects the fact that
over a $4\pi$ circuit we visit the axis of each ellipse on $\sigma_{0}$ twice,
arriving at opposite ends of the axis after a $2\pi$ circuit. \ From this
inversion symmetry it follows that after any $2\pi$ circuit the tangent
vectors at the beginning and end of the circuit must be antiparallel, so that
$n$ must be an odd positive/negative integer.%

\begin{figure}
[h]
\includegraphics[width=0.5\textwidth]%
{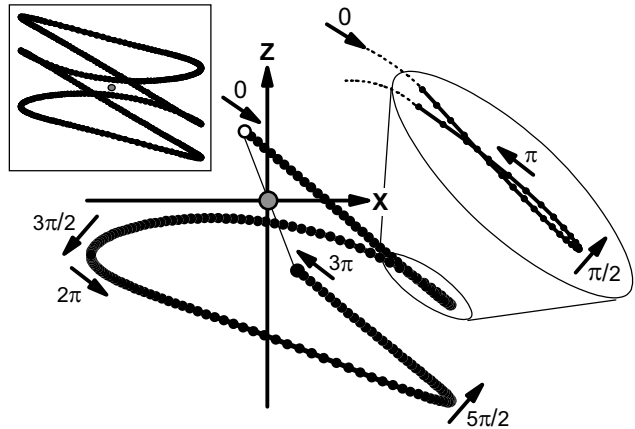}%
\caption{Tangent index $d\tau_{\alpha,\beta}$. \ As in Figs. \ref{Fig8} and
\ref{Fig9}a, the central figure shows the curve (black dots) generated by the
projection onto $\tau_{0}$ of the endpoints of axis $\boldsymbol{\alpha}$ of
the ellipses on the surrounding circle $\sigma_{0}$. \ The point where
$\sigma_{0}$ pierces $\tau_{0}$ is shown by the gray circle. \ The starting
point of a $2\pi$ circuit on $\sigma_{0}$ is shown by the white dot, the end
of the circuit by the black dot that terminates the endpoint curve. \ The
tangent vector to the endpoint curve is shown by arrows labelled with the
orientation of the vector relative to the starting orientation, $0$. \ As can
be seen, the vector rotates through $+3\pi$ over the circuit, so that
$d\tau_{\alpha,\beta}$, which measures the net rotation of the vector, is
$d\tau_{\alpha,\beta}=+3/2$. The inset in the upper left corner shows the
endpoint\ curve for a $4\pi$ circuit. \ As can be seen, this curve is
centrosymmetric about $\sigma_{0}$. \ $\tau_{\alpha,\beta}$ can be determined
by inspection using the phase ratchet rules. Here the sequence of crossings of
the curve with the $XZ$-axes is $Z_{-}X_{-}Z_{-}Z_{+}$; after applying the
rules this reduces to $Z_{-}X_{-}$, showing that $\tau_{\alpha,\beta}=-1/2$. }%
\label{Fig10}%
\end{figure}

\subsubsection{Index $d\tau_{\gamma}$}

As shown in Fig. \ref{Fig7}, axis $\boldsymbol{\gamma}$ does not generate a
M\"{o}bius strip. \ It may therefore appear surprising that as $\tau_{0}$
rotates along $\sigma_{0\text{ }}$ the projections onto this plane of the
$\boldsymbol{\gamma}$-axis endpoints generate a two turn helix. This helix,
which can be either right- or left-handed, is characterized by tangential
winding number $d\tau_{\gamma},$ the equivalent for axis $\boldsymbol{\gamma}$
of the index $d\tau_{\alpha,\beta}$ for axes $\boldsymbol{\alpha}$ and
$\boldsymbol{\beta}$. \ As can be seen in Fig. \ref{Fig7}, the endpoints of
axis $\boldsymbol{\gamma}$ return to themselves after one $2\pi$ circuit, so
$d\tau_{\gamma}=\pm n$; we find in all cases $\left\vert n\right\vert =2$,
reflecting the fact that both turns of the helix are identical. \ Like a
M\"{o}bius strip, a helix retains its handedness, and therefore its winding
number, when viewed from any direction. \ Examples of $d\tau_{\gamma}$ are
shown in Fig. \ref{Fig11}, which also shows that the index $\tau
_{\boldsymbol{\gamma}}$, the \textbf{$\boldsymbol{\gamma}$} equivalent of
$\tau_{\alpha,\beta}$, is always zero, and is therefore of no interest.%

\begin{figure}
[h]
\includegraphics[width=0.5\textwidth]%
{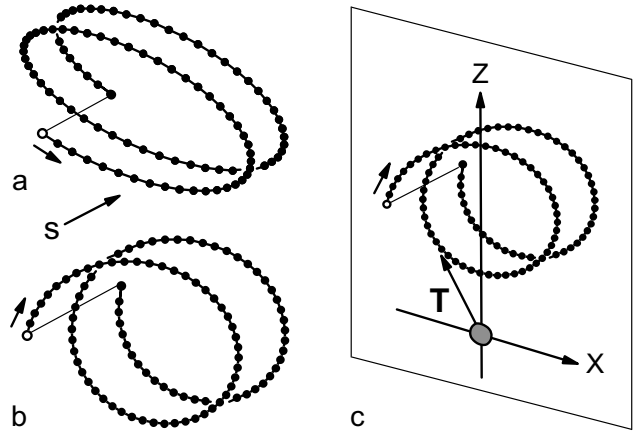}%
\caption{Tangent index $d\tau_{\gamma}$. \ Axis $\boldsymbol{\gamma}$
$\ $endpoints (black dots) are plotted as a function of arclength $s=r_{0}%
\chi$, where $\chi$ measures position on $\sigma_{0}$, Fig. \ref{Fig3}a, and
$r_{0}$ is the radius of $\sigma_{0}$, \ (a) The endpoints generate a
left-handed helix. \ Starting at the white dot, the tangent vector to the
curve (arrow) rotates monotonically in the positive, counterclockwise
direction through $4\pi$, and $d\tau_{\gamma}=+2$. \ (b) Here the endpoints
form a right-handed helix, the tangent vector rotates through $4\pi$ in the
negative, clockwise direction, and $d\tau_{\gamma}=-2$. \ (c) Index
$\tau_{\gamma}$. \ Vector $\mathbf{T}$ oscillates back and forth with a net
winding angle of zero. \ Starting at the white dot and proceeding clockwise,
the crossing sequence of the endpoint curve with the $XZ$-axes is $Z_{-}%
Z_{+}Z_{-}Z_{+}$. \ Applying the rules of the phase ratchet yields an empty
sequence, so $\tau_{\gamma}=0$; this is the value of $\tau_{\gamma}$ for all
$\boldsymbol{\gamma}$ cones. \ Unlike $\boldsymbol{\alpha}$ and
$\boldsymbol{\beta}$, whose endpoint curves are typically complex, Figs.
\ref{Fig9} and \ref{Fig10}, $\boldsymbol{\gamma}$ endpoint helices always have
smooth, simple shapes. }%
\label{Fig11}%
\end{figure}

\subsection{Topological Indices of the Projection onto ${\protect\Large \pi
}_{0}$: Indices $\pi_{\alpha,\beta}$, $d\pi_{\alpha,\beta}$, and d$\pi
_{\gamma}$}

The projections of the axes of the ellipses on $\sigma_{0}$ onto the plane
$\pi_{0}$ in Fig. \ref{Fig3}b generate three winding numbers that are analogs
of $\tau_{\alpha,\beta}$, $d\tau_{\alpha,\beta}$, and $d\tau_{\gamma}$.

The $\pi_{0}$ analog of $\tau_{\alpha,\beta}$ we label $\pi_{\alpha,\beta}$;
like $\tau_{\alpha,\beta}$, $\pi_{\alpha,\beta}$ takes on the values
$\pm1/2,\pm3/2$, is the same\ for axes $\boldsymbol{\alpha}$ and
$\boldsymbol{\beta}$ in $\Sigma_{0}$, and can be determined by inspection
using the phase ratchet rules with axes $XZ$ replaced by\ $YZ$. \ All $16$
combinations of $\tau_{\alpha,\beta}$ and $\pi_{\alpha,\beta}$ appear in our
simulations, showing that these two indices are independent one of the other.

The $\pi_{0}$ analog of $d\tau_{\alpha,\beta}$ we label $d\pi_{\alpha,\beta}$;
like $d\tau_{\alpha,\beta}$, $d\pi_{\alpha,\beta}$ takes on the values
$\pm1/2,\pm3/2$, is the same for axes $\boldsymbol{\alpha}$ and
$\boldsymbol{\beta}$ in $\Sigma_{0}$, and can be determined by following the
rotation of the tangent vector to the endpoint curve in $\pi_{0}$. \ We show
in Section IV that, just like for $\tau_{\alpha,\beta}$ and $d\tau
_{\alpha,\beta}$, only ten of the sixteen possible combinations of
$\pi_{\alpha,\beta}$ and $d\pi_{\alpha,\beta}$ appear, so that these two
indices are not completely independent one of the other.

The $\pi_{0}$ analog of $d\tau_{\gamma}$ we label $d\pi_{\gamma}$; like
$d\tau_{\gamma}$, $d\pi_{\gamma}$ takes on the values $\pm2$, and can be
determined by following the rotation of the tangent vector to the
$\boldsymbol{\gamma}$ endpoint curve. \ All $4$ combinations of $d\tau
_{\gamma}$ and $d\pi_{\gamma}$ appear in our simulations, showing that these
two indices are independent one of the other.

$\pi_{\alpha,\beta}$, $d\pi_{\alpha,\beta}$, and $d\pi_{\gamma}$ are
illustrated in Fig. \ref{Fig12}.%

\begin{figure}
[h]
\includegraphics[width=0.65\textwidth]%
{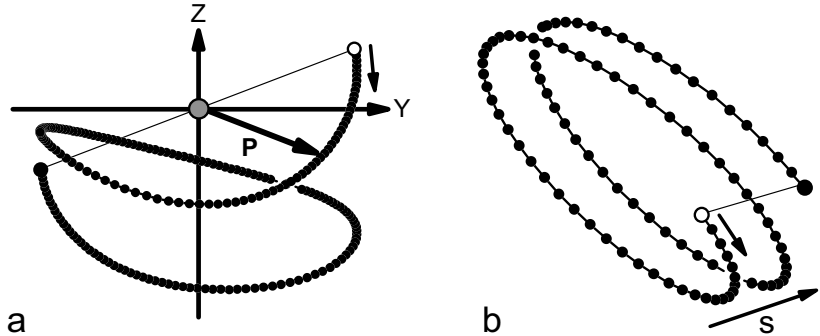}%
\caption{Indices $\pi_{\alpha,\beta}$, $d\pi_{\alpha,\beta}$, and
$d\pi_{\gamma}$. \ (a) $\pi_{\alpha,\beta}$ and $d\pi_{\alpha,\beta}$.
\ $\pi_{\alpha,\beta}$ measures the net rotation of vector $\mathbf{P}$.
\ Starting at the small white circle, the crossing sequence of the endpoint
curve (small black dots) with the $YZ$-axes is $Y_{-}Z_{-}Z_{+}Z_{-}$, which,
after application of the phase ratchet rules reduces to $Y_{-}Z_{-}$, showing
that $\pi_{\alpha,\beta}=-1/2$. \ $d\pi_{\alpha,\beta}$ measures the net
rotation of the tangent vector (small arrow) to the endpoint curve. \ This
vector is easily seen to rotate by $3\pi$ in the negative, clockwise
direction, and $d\pi_{\alpha,\beta}=-3/2$. \ (b) $d\pi_{\gamma}$.
$d\pi_{\gamma}$ measures the net rotation of the tangent vector (small arrow)
to the $\gamma$ helix, plotted here, as in Fig. 11, vs. arclength $s$ on
$\sigma_{0}$. \ As can be seen, the tangent vector rotates in the negative,
clockwise direction by $4\pi$, and $d\pi_{\gamma}=-2$. \ The data in (a) (in
(b)) for $\pi_{\alpha,\beta}$ and $d\pi_{\alpha,\beta}$ (for $d\pi_{\gamma}$)
are for the same M\"{o}bius strip (same $\mathcal{\gamma}$ cone) as that shown
in Fig. \ref{Fig10} (in Fig. \ref{Fig11}a). \ }%
\label{Fig12}%
\end{figure}

\subsection{Line Classification}

In addition to determining winding numbers $I_{\alpha,\beta}$ for axes
$\boldsymbol{\alpha}$ and $\boldsymbol{\beta}$ (Section III.A.1), and
$I_{\gamma}$ for axis $\boldsymbol{\gamma}$ (Section III.A.2), the projections
of these axes onto $\Sigma_{0}$ determines another important property of
ellipse fields: the line classification. \ In this classification, introduced
by Berry and Hannay [$18$], one counts the number of straight streamlines that
terminate (or equivalently originate) on a singularity;\ here we denote this
number by $\Lambda$. $\ $We discuss this classification for C points for axes
$\boldsymbol{\alpha}$ and $\boldsymbol{\beta}$, and we then extend it to axis
$\boldsymbol{\gamma}$.

\subsubsection{$\Lambda_{\alpha,\beta}$}

We start be reviewing well known results about application of the line
classification to C points. \ Berry and Hannay showed that within the linear
approximation, for umbilic points there are only two possibilities $\Lambda=1$
and $\Lambda=3$ [$18$]. \ At an umbilic point the eigenvalues of the curvature
tensor (matrix) become degenerate. \ This tensor can be represented by an
ellipse, and at an umbilic point the major and minor axes of the ellipse
become equal and the ellipse degenerates to a circle. \ From this it easily
follows that the results of Berry and Hannay are applicable to C points, which
as discussed in Section II.A, are points where the eigenvalues of the
coherency matrix become degenerate and the polarization ellipse degenerates
into a circle $-$ the C circle.

When the path that encircles the C point is a circle with sufficiently small
radius, $\Lambda$ is the number of lines in plots such as those in Figs.
\ref{Fig2}, \ref{Fig4}, and \ref{Fig6} that are radially directed towards the
C point. \ Labelling $\Lambda$ by the relevant axis we have from Berry and
Hannay, $\Lambda_{\alpha}=\Lambda_{\beta}=\Lambda_{\alpha,\beta}=1,3$.

The reason why $\Lambda_{\alpha}=\Lambda_{\beta}$ for a C point is as follows:
\ An important geometric property of the lines in Figs. 2, 4, and 6, is that
lines on opposite sides of $\sigma_{0}$ are orthogonal. \ Thus, for every line
that points towards the C point there is a line on the opposite side of the
circle that is tangent to $\sigma_{0}$. \ But as discussed in Section II.A.1,
for $\sigma_{0}$ sufficiently small the lines representing the projections
onto $\Sigma_{0}$ of axes $\boldsymbol{\alpha}$ and $\boldsymbol{\beta}$ are
also orthogonal, so that wherever $\boldsymbol{\alpha}$ is tangent to
$\sigma_{0}$, $\boldsymbol{\beta}$ points towards the C point, and vice versa.

Here we determine $\Lambda_{\alpha,\beta}$ numerically, finding the number of
points on $\sigma_{0}$ at which the orientation $\theta$ of the ellipse axis,
measured counterclockwise relative to the $x$-axis, equals $\chi$, the angular
position of the axis center on $\sigma_{0}$, Fig. \ref{Fig3}. \ Specifically,
calculating $\theta$ for say axis $\boldsymbol{\alpha}$ as $\theta
=\arctan\left(  \alpha_{y},\alpha_{x}\right)  $, where $\alpha_{x}$
($\alpha_{y}$) are the components (projections) of $\boldsymbol{\alpha}$ along
the $x$- and $y$-axes, respectively, we solve $\theta\left(  \chi\right)
=\chi+n\pi$, where $n=-4$ $...$ $+4$, as required [$32$, p. 258]. \ Fig.
\ref{Fig13}a illustrates a graphical implementation of this method. \ We note
that an alternative, analytical, method would be to use the discriminant given
by Dennis [$33$], writing the Stokes parameters in $\Sigma_{0}$ in terms of
$\alpha_{x}$ and $\alpha_{y}$ [$13$], but the resulting expressions for the
general case are so unwieldy as to be impractical.

There is an important connection between $I_{\alpha,\beta}$ and $\Lambda
_{\alpha,\beta}$. \ Because $I_{\alpha,\beta}=+1/2,-1/2$ and $\Lambda
_{\alpha,\beta}=1,3$, it might be thought that there are four possible
combinations of these indices. Berry and Hannay showed, however, that this is
not the case and that the number of possible combinations is only three:
\ (\emph{i}) $I_{\alpha,\beta}=+1/2$, $\Lambda_{\alpha,\beta}=1$, the
\emph{lemon}; (\emph{ii}) $I_{\alpha,\beta}=+1/2$, $\Lambda_{\alpha,\beta}=3$,
the \emph{monstar}; and (\emph{iii}) $I_{\alpha,\beta}=-1/2$, $\Lambda
_{\alpha,\beta}=3$, the \emph{star}. \ These three combinations are
illustrated in Fig. \ref{Fig13} b $-$ f.%

\begin{figure}
[h]
\includegraphics[width=0.9\textwidth]%
{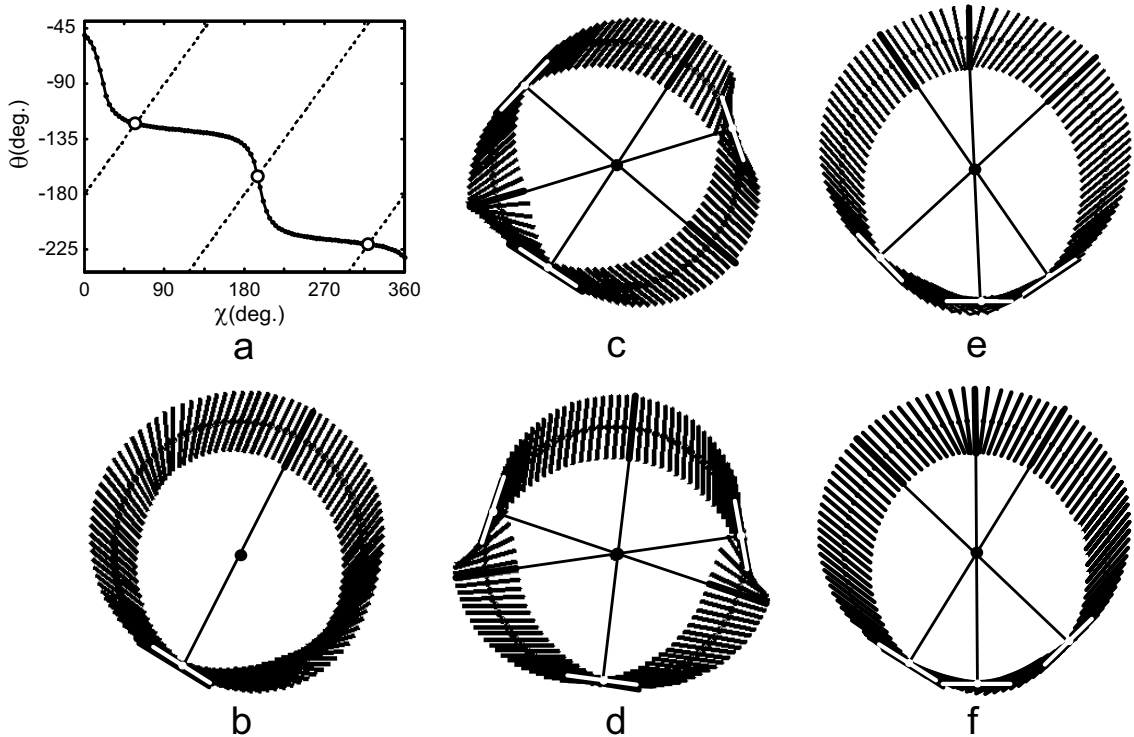}%
\caption{$\Lambda_{\alpha,\beta}$. \ (a) Graphic implementation of the
numerical method described in the text for determining $\Lambda$. \ The solid
curve is the angle $\theta\left(  \chi\right)  $ that the projections of axes
$\boldsymbol{\alpha}$ make with the $x$-axis. \ $\Lambda$ counts the
intersections (small white circles) of this curve with the dotted lines
$\theta\left(  \chi\right)  =$ $\chi+n\pi$, where here $n=-1$, $-2$, and $-3$.
\ These data are for the projection shown here in (c). \ (b) A $\left\vert
\tau_{\alpha,\beta}\right\vert =1/2$ lemon. Here $I_{\alpha,\beta}=+1$,
$\Lambda_{\alpha,\beta}=1$, and $\tau_{\alpha,\beta}=+1/2$ . \ Throughout, a
thick line marks an axis that points directly at the C point (central dot);
perpendicular to this is an axis, marked by a white line, that is tangent to
the $\sigma_{0}$ circle. \ (c) A $\left\vert \tau_{\alpha,\beta}\right\vert
=1/2$ star. Here $I_{\alpha,\beta}=-1$, $\Lambda_{\alpha,\beta}=3$, and
$\tau_{\alpha,\beta}=-1/2$. \ (d) A $\left\vert \tau_{\alpha,\beta}\right\vert
=3/2$ star. Here $I_{\alpha,\beta}=-1$, $\Lambda_{\alpha,\beta}=3$, and
$\tau_{\alpha,\beta}=+3/2$. \ (e) A $\left\vert \tau_{\alpha,\beta}\right\vert
=1/2$ monstar. $\ $Here $I_{\alpha,\beta}=+1$, $\Lambda_{\alpha,\beta}=3$, and
$\tau_{\alpha,\beta}=-1/2$. \ (f) A $\left\vert \tau_{\alpha,\beta}\right\vert
=3/2$ monstar. $\ $Here $I_{\alpha,\beta}=+1$, $\Lambda_{\alpha,\beta}=3$, and
$\tau_{\alpha,\beta}=+3/2$. \ As discussed in the text, there are no
$\left\vert \tau_{\alpha,\beta}\right\vert =3/2$ lemons.}%
\label{Fig13}%
\end{figure}

There are also important connections between $\Lambda_{\alpha,\beta}$, and
$\tau_{\alpha,\beta}$, $\pi_{\alpha,\beta}$, the phase ratchet rules, and the
endpoint curves in Figs. \ref{Fig8}, \ref{Fig9}, and \ref{Fig10}; these
connections follow from the facts that when the endpoint curve of say
$\boldsymbol{\alpha}$ crosses the $Z$-axis of the plane $\tau_{0}$, the
projection of $\boldsymbol{\alpha}$ onto $\Sigma_{0}$ is \emph{tangent} to
$\sigma_{0}$, whereas when the endpoint curve crosses the $Z$-axis of $\pi
_{0}$, the projection of $\boldsymbol{\alpha}$ onto $\Sigma_{0}$ is
\emph{radial} (along the radius of $\sigma_{0}$). \ These important
geometrical results are illustrated in Fig. \ref{Fig14}.%

\begin{figure}
[h]
\includegraphics[width=0.7\textwidth]%
{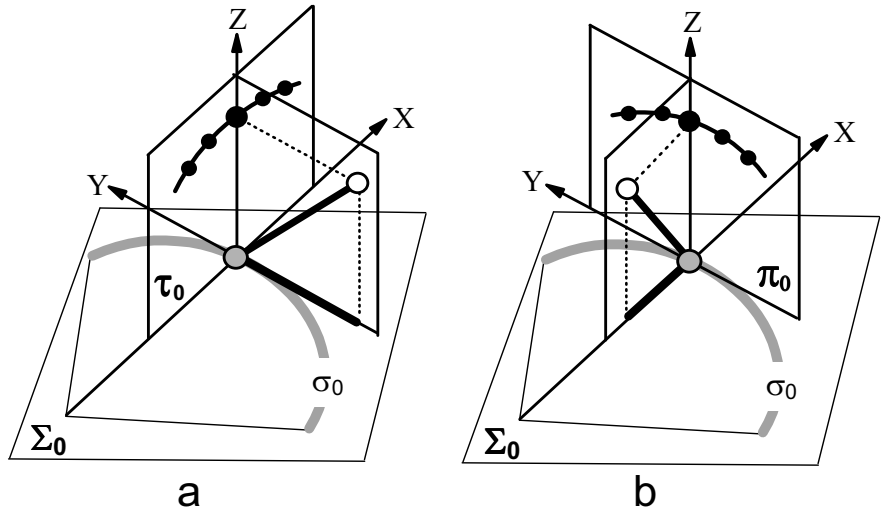}%
\caption{$Z$-axis crossings. \ Ellipse axis centers (endpoints) are shown by
small gray (white) circles, whereas the axis itself is shown by the connecting
thick line. \ The projected endpoint curves in the planes $\tau_{0}$ (the
$XZ$-plane) and $\pi_{0}$ (the $YZ$-plane) are shown by the connected black
dots. (a) When the endpoint curve in $\tau_{0}$ crosses the $Z$-axis, the
ellipse axis lies in the $YZ$-plane, and its projection onto $\Sigma_{0}$
is\emph{ tangent} to the $\sigma_{0}$ circle$.$ (b) When the endpoint curve in
$\pi_{0}$ crosses the $Z$-axis, the ellipse axis lies in the $XZ$-plane, and
its projection onto $\Sigma_{0}$ is\emph{ radial}. }%
\label{Fig14}%
\end{figure}

We therefore have the following:

\bigskip

(i) Endpoint crossings with the $Z$-axis in $\tau_{0}$ and the $Z$-axis in
$\pi_{0}$ occur at opposing points on $\sigma_{0}$ (points separated by
$180^{o}$).

(ii) $\Lambda_{\alpha,\beta}$ equals the number of endpoint crossings of the
$Z$-axis in $\tau_{0}$: at each crossing the axis projection onto $\Sigma_{0}$
is tangent to $\sigma_{0}$, implying the presence of a radial projection at
the opposing point on $\sigma_{0}$; by definition, $\Lambda_{\alpha,\beta}$ is
the number of such radial projections.

($iii$) $\Lambda_{\alpha,\beta}$ also equals the number of endpoint crossings
of the $Z$-axis in $\pi_{0}$, because at each such crossing the axis
projection onto $\Sigma_{0}$ is radial.

\bigskip

From (ii) and (iii) follows that the number of endpoint crossings with the
$Z$-axis is the same for both $\tau_{0}$ and $\pi_{0}$, and that this number
must be either $1$ or $3$. \ Below, for simplicity we refer to an endpoint
crossing with the $Z$-axis as a \textquotedblleft crossing\textquotedblright,
and we have the following:

\bigskip

(\emph{I}) If $\Lambda_{\alpha,\beta}=1$, $\left\vert \tau_{\alpha,\beta
}\right\vert =1/2$: from the phase ratchet rules if $\left\vert \tau
_{\alpha,\beta}\right\vert =3/2$ there must be more than one crossing. \ Thus,
all lemons are M\"{o}bius strips with a single half-twist. \ The converse,
however, is not true, and a one-half-twist strip can have $\Lambda
_{\alpha,\beta}=3$, and can therefore be a star or a monstar, Figs \ref{Fig13}c,d

(\emph{II}) If $\left\vert \tau_{\alpha,\beta}\right\vert =3/2$, then
$\Lambda_{\alpha,\beta}=3$, so all M\"{o}bius strips with three half-twists
are either stars or monstars. \ This important result is to be credited to Dr.
Mark R. Dennis (Bristol), who, using the methods of differential geometry, was
first to derive it [$34$].

(\emph{III}) Within the linear approximation the M\"{o}bius strips surrounding
C lines can have one or three half-twists $-$ strips with say five half-twists
cannot occur within this approximation. \ Strips not necessarily associated
with C lines, or with L lines (which have four half-twists), are also
possible; an example is the two-half-twist (one-full-twist) M\"{o}bius strips
that surround ordinary ellipses [$14,15$].

\subsubsection{$\Lambda_{\gamma}$}

For the line classification for axis $\boldsymbol{\gamma}$ we find the
following: \ If $I_{\gamma}=+1,$ $\Lambda_{\gamma}=0,4$, whereas if
$I_{\gamma}=-1$, $\Lambda_{\gamma}=4$; examples of all three cases are shown
in Fig. \ref{Fig6}, where, as in Fig. \ref{Fig13} for $\Lambda_{\alpha,\beta}%
$, axes that point toward the C point are marked by thick lines. \ Unlike the
case of axes $\boldsymbol{\alpha}$ and $\boldsymbol{\beta}$, however, at
opposing points on $\sigma_{0}$, $\boldsymbol{\gamma}$ axes are not
perpendicular, but are parallel. \ Whether $\Lambda_{\gamma}=0$, or
$\Lambda_{\gamma}=4$, depends on whether the endpoint curve in $\pi_{0}$
crosses the $Z$-axis, because the geometrical considerations shown in Fig.
\ref{Fig14}b hold for all three axes $-$ $\boldsymbol{\alpha}$,
$\boldsymbol{\beta}$, and $\boldsymbol{\gamma}$. \ In Fig. \ref{Fig15} we show
the three endpoint curves that correspond to the three cases in Fig.
\ref{Fig6}. \ As can be seen, when $\Lambda_{\gamma}=0$ ( $\Lambda_{\gamma}%
=4$) the endpoint curve crosses the $Z$-axis $0$ ($4$) times.%

\begin{figure}
[h]
\includegraphics[width=0.75\textwidth]%
{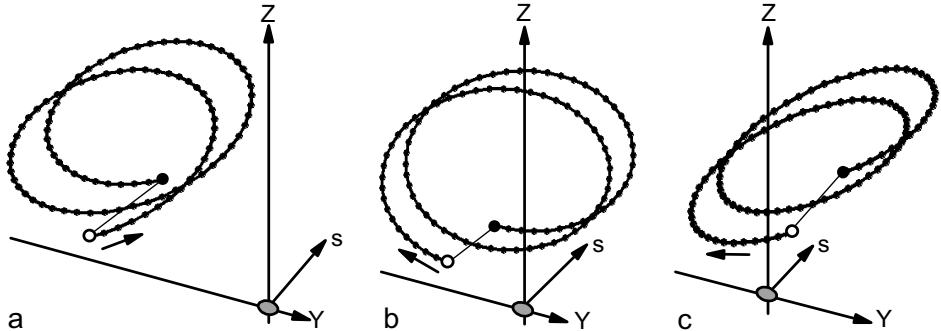}%
\caption{Axis $\boldsymbol{\gamma}$ endpoint curves in $\pi_{0}$. \ Panels
(a), (b), and (c), are the endpoint curves corresponding to panels (a), (b),
and (c) in Fig. \ref{Fig6}, respectively. \ As can be seen, here in (a) the
endpoint curve does not cross the $Z$-axis and in (a) of Fig. \ref{Fig6}
$\Lambda_{\gamma}=0$ , as expected. \ In (b) and (c) there are four crossings
and in (b) and (c) of Fig. \ref{Fig6} $\Lambda_{\gamma}=4$, again as
expected.}%
\label{Fig15}%
\end{figure}

\subsection{Index Summary}

We summarize here the rather complex results discussed above for the various
indices. \ All indices are the same for axes $\boldsymbol{\alpha}$ and
$\boldsymbol{\beta}$ and are therefore labelled \textquotedblleft$\alpha
,\beta$\textquotedblright.

(\emph{i}) Index $I_{\alpha,\beta}$ measures the rotation of the projections
of axes $\boldsymbol{\alpha}$ and $\boldsymbol{\beta}$ onto $\Sigma_{0}$, the
plane of the C circle, Fig. \ref{Fig2}. \ Generically $I_{\alpha,\beta}%
=\pm1/2$. \ A positive (negative) value for this index, and for index
$I_{\gamma}$ below, implies that the axis projections rotate in the same
(opposite or retrograde) direction as the path on which they are measured.

(\emph{ii}) There are either one or three points on a sufficiently small
circular path centered on the C point at which the axis projections point
directly towards the point. \ The number of such points is measured by the
line classification index $\Lambda_{\alpha,\beta}$. Unlike $I_{\alpha,\beta}$,
$\Lambda_{\alpha,\beta}$ is not a topological invariant, but is nonetheless an
important, highly useful characterization.

(\emph{iii}) If $I_{\alpha,\beta}=+1/2$ and $\Lambda_{\alpha,\beta}=1$
($\Lambda_{\alpha,\beta}=3$), the C point is dubbed a \textquotedblleft
lemon\textquotedblright\ (a \textquotedblleft monstar\textquotedblright), Fig.
\ref{Fig13}. \ Lemons always correspond to M\"{o}bius strips with a single
half-twist, whereas monstars can be M\"{o}bius strips with one or three
half-twists. \ If $I_{\alpha,\beta}=-1/2$, then $\Lambda_{\alpha,\beta}=3$,
and the C point is dubbed a \textquotedblleft star\textquotedblright. \ Stars
can correspond to M\"{o}bius strips with one or three half-twists.
\ Conversely, if a M\"{o}bius strip has a single half-twist without any
mutually cancelling additional twists (Fig. \ref{Fig9}b) it must be a lemon,
whereas if it has three half-twists it can be either a star or a monstar.
\ The above relationships are the same for both right- and left-handed
M\"{o}bius strips.

(\emph{iv}) Index $I_{\gamma}$ measures the rotation of the projection of axis
$\boldsymbol{\gamma}$ onto $\Sigma_{0}$; generically $I_{\gamma}=\pm1$, Fig.
\ref{Fig6}. \ The line classification index $\Lambda_{\gamma}$ measures the
number of points on the path at which $\boldsymbol{\gamma}$ axis projections
point directly towards the C point. \ If $I_{\gamma}=+1$, $\Lambda_{\gamma
}=0,4$, whereas for $I_{\gamma}=-1$, $\Lambda_{\gamma}=4$. \ $I_{\alpha,\beta
}$ and $I_{\gamma}$ are independent, and all four combinations of these
indices can appear. \ $I_{\gamma}$ and $\Lambda_{\gamma}$ are independent of
the number of twists or handedness of the M\"{o}bius strip.

(\emph{v}) The magnitude of index $\tau_{\alpha,\beta}$ measures the net
number of half twists of the M\"{o}bius strip, the sign of $\tau_{\alpha
,\beta}$ measures the handedness of the strip ($+=$ left-handed, $-=$
right-handed), Figs. \ref{Fig8} and \ref{Fig9}. \ There are four possible
values for $\tau_{\alpha,\beta}$: $\tau_{\alpha,\beta}=+1/2,-1/2,+3/2,-3/2$.

(\emph{vi}) There are in principle $2\times2\times4=16$ possible triple
combinations of $I_{\alpha,\beta}$, $\Lambda_{\alpha,\beta}$, and
$\tau_{\alpha,\beta}$, but (see (\emph{iii}) above) only $10$ are allowed:
$I_{\alpha,\beta}=+1/2$, $\Lambda_{\alpha,\beta}=1$, $\left\vert \tau
_{\alpha,\beta}\right\vert =1/2$ (a one-half-twist lemon); $I_{\alpha,\beta
}=+1/2$, $\Lambda_{\alpha,\beta}=3$, $\left\vert \tau_{\alpha,\beta
}\right\vert =1/2,3/2$ (a one- and a three-half-twist monstar); $I_{\alpha
,\beta}=-1/2$, $\Lambda_{\alpha,\beta}=3$, $\left\vert \tau_{\alpha,\beta
}\right\vert =1/2,3/2$ (a one- and a three-half-twist star).

(\emph{vii}) $\tau_{\alpha,\beta}$ is independent of $I_{\gamma}$ and
$\Lambda_{\gamma}$, and for each of the three allowed $I_{\gamma}$ and
$\Lambda_{\gamma}$ combinations in (\emph{iv}) above, all four values of
$\tau_{\alpha,\beta}$ appear, so that in total there are $12$ triple
combinations of $I_{\gamma}$, $\Lambda_{\gamma}$, and $\tau_{\alpha,\beta}$.

(\emph{viii}) Whereas $\tau_{\alpha,\beta}$ measures the net signed number of
times axes $\boldsymbol{\alpha}$ and $\boldsymbol{\beta}$ rotate or loop
around the line of the path, index $d\tau_{\alpha,\beta}$ measures additional
signed rotations or loops that do not encircle this line, Fig. \ref{Fig10}.
\ $d\tau_{\alpha,\beta}$ takes on the four values $d\tau_{\alpha,\beta}%
=\pm1/2,\pm3/2$. \ A right-handed one-half-twist M\"{o}bius strip
($\tau_{\alpha,\beta}=-1/2$), for example, can have $d\tau_{\alpha,\beta
}=+3/2$ due to additional loops that are \textquotedblleft up in the
air\textquotedblright, as it were.

(\emph{ix}) Indices $\pi_{\alpha,\beta}$ and $d\pi_{\alpha,\beta}$ measure
oscillations of axes $\boldsymbol{\alpha}$ and $\boldsymbol{\beta}$ \ that
form signed loops not captured by indices $\tau_{\alpha,\beta}$ and
$d\tau_{\alpha,\beta}$:\ $\pi_{\alpha,\beta}=\pm1/2,\pm3/2$; $d\pi
_{\alpha,\beta}=\pm1/2,\pm3/2$, Fig. \ref{Fig12}. \ $\pi_{\alpha,\beta}$ is
independent of $\tau_{\alpha,\beta}$, $d\pi_{\alpha,\beta}$ is independent of
$d\tau_{\alpha,\beta}$.

(\emph{x}) Indices $d\tau_{\gamma}$ and $d\pi_{\gamma}$ measure
\textquotedblleft up in the air\textquotedblright\ oscillations of axis
$\boldsymbol{\gamma}$ that form signed loops which do not enclose the line of
the path, Fig. \ref{Fig11}. \ These indices describe the cones generated by
axis $\boldsymbol{\gamma}$, Fig. \ref{Fig7}, are independent of one another,
and of the corresponding $\boldsymbol{\alpha},\boldsymbol{\beta}$ indices, and
take the values: $d\tau_{\gamma}=\pm2$; $d\pi_{\gamma}=\pm2$.

Altogether there are a total of ten indices that characterize C points, their
M\"{o}bius strips, and their cones: $I_{\alpha,\beta},I_{\gamma}%
,\Lambda_{\alpha,\beta},\Lambda_{\gamma},\tau_{\alpha,\beta},d\tau
_{\alpha,\beta},\pi_{\alpha,\beta},d\pi_{\alpha,\beta},d\tau_{\gamma}%
,d\pi_{\gamma}$. \ If independent, these indices could generate $2^{4}%
\times4^{4}\times2^{2}=16384$ different C points (C lines). \ The connections
discussed above reduce this number to $7680$. \ In Section IV we discuss
additional connections that reduce this number even further, and present
statistical data for the various index combinations.

\section{STATISTICS}

Here we list the most important statistical properties of the geometrically and topologically distinct Mobius strips that appear in $\Sigma_{0}$. \ We emphasize that these statistics change importantly when the plane of observation $\Sigma$ is
rotated away from $\Sigma_{0}$ by small, but finite angles.

As noted in Section III.E, the ten indices $I_{\alpha,\beta},I_{\gamma
},\Lambda_{\alpha,\beta},\Lambda_{\gamma},\tau_{\alpha,\beta},d\tau
_{\alpha,\beta},\pi_{\alpha,\beta},d\pi_{\alpha,\beta},d\tau_{\gamma}%
,d\pi_{\gamma}$, if independent, could generate $16384$ different C points (C
lines). \ This number is reduced to $7680$ by the constraints discussed in
Section III.D. \ In this section we search for additional constraints
(selection rules) that further reduce the number of allowed index
combinations, and we present statistical data for the probabilities of the
remaining combinations.

We search a database containing $10^{6}$ independent, randomly generated
realization, after removing all realizations in which any one of the ten
indices differed by more than $\pm0.001$ from an integer or half integer.
\ This eliminated $\sim2\%$ of the realizations. \ The the remaining index
combinations, which all obey the constraints discussed in Section III.D, have
a very broad distribution. \ This is illustrated in Fig. \ref{Fig21} using a
modified Zipf plot. \ As can be seen from this plot, it is unlikely that the
number of configurations appearing in an arbitrarily large database would
substantially exceed $1,200$.%

\begin{figure}
[h]
\includegraphics[width=0.65\textwidth]%
{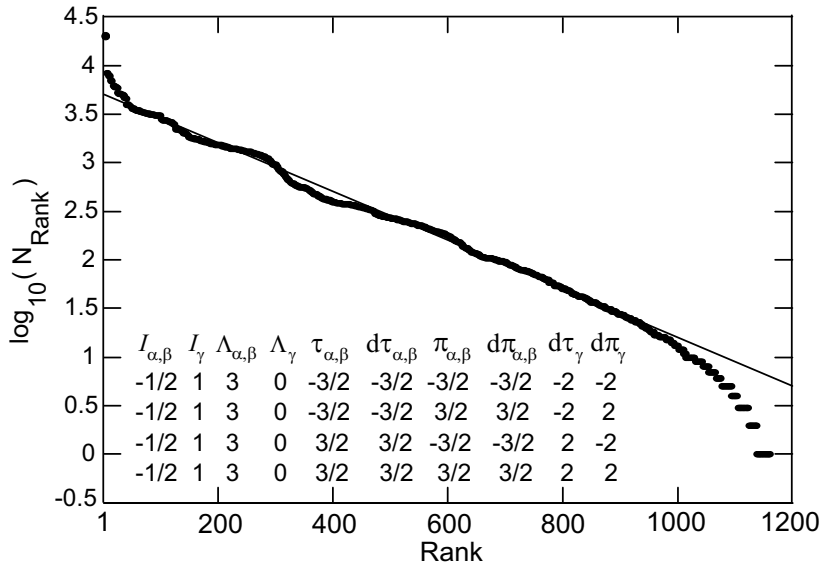}%
\caption{Modified Zipf plot. \ The logarithm of the number of exemplars,
$N_{Rank}$, found for a given ten-index configuration is plotted vs. the rank
of the configuration, where configurations are ranked in order of decreasing
probability. \ The straight line is $N_{Rank\text{ }}=N_{o}\exp\left(
-K\times Rank\right)  $, with $N_{o}=4950$ and $K=0.00575$. \ The tabular
inset lists the four most probable configurations (the four unresolved
points\ in the upper left corner). \ There are some $2\times10^{4}$ exemplars
of each of these four configurations in the complete $10^{6}$ entry database.
\ As can be seen, these four most probable configurations are all
three-half-twist stars.}%
\label{Fig21}%
\end{figure}

\subsection{Selection Rules}

It is useful to formulate constraints among indices in the form of selection
rules $-$ rules that state which combinations of indices are forbidden. \ Here
we list $13$ such rules: $8$ rules involve pairs of indices $-$\ \emph{binary
rules}; $5$ rules involve index triplets \ $-$\ \emph{ternary rules}.

The search for binary (ternary) rules is facilitated by 2D (3D) plots such as
those shown in Fig. \ref{Fig22}. \ There are, in principal, a total of $45$ 2D
plots and $120$ 3D plots that need to be examined and interpreted. \ Our
algorithms, however, display for analysis only those plots in which there are
missing indices, reducing the number of plots to $8$ for 2D, and $62$ for 3D.
\ Of the 3D plots, $57$ do not yield new ternary rules, because the absence of
configurations results from combinations of the binary rules. \ An example of
this is shown in Fig. \ref{Fig22}c.%

Altogether, the $13$ selection rules reduce the number of index configurations
from $16,384$ to $2,104$. \ This is almost twice the number observed, Fig.
\ref{Fig21}, indicating the possible presence of additional, higher-order
rules. \ There is no simple way to systematically search for such rules,
however, since in the overwhelming majority of cases the absence of index
configurations for, say the $210$ quaternary combinations, will be due to
binary and ternary selection rules acting together. \ Because of this we have
not attempted higher-order searches.

\begin{figure}
[h]
\includegraphics[width=0.95\textwidth]%
{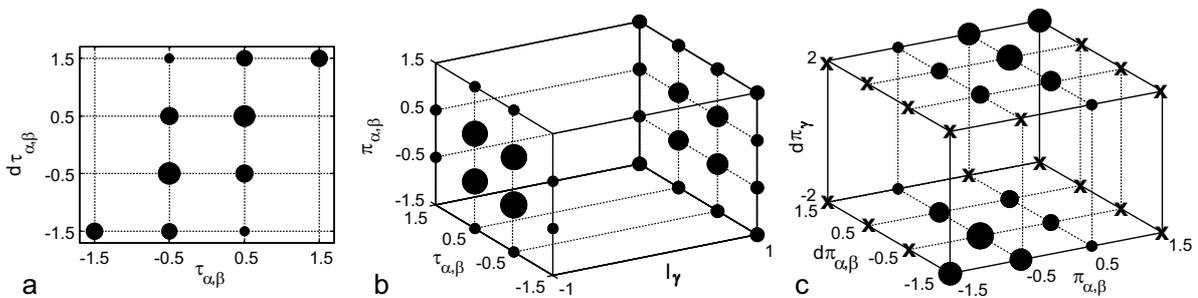}%
\caption{Index combinations. \ (a) 2D plot for binary index combinations.
\ (b),(c) 3D plots for ternary combinations. \ In each case empty grid points
correspond to missing, and therefore presumed forbidden, combinations. \ The
areas of the black circles are approximately proportional to the relative
probabilities of the different combinations. \ (a) Binary selection Rule 2.5
is derived from this plot. \ (b) Plot from which ternary selection Rule 3.4 is
derived. \ (c) Here all missing indices are due to combinations of binary
rules. \ The interested reader may enjoy analyzing the $16$ configurations
(marked by x's) missing from this plot in terms of the rules.\ \ }%
\label{Fig22}%
\end{figure}

\subsubsection{Binary Rules}

Listed below are the $8$ binary selection rules (forbidden index
combinations). The number of configurations each rule eliminates when acting
alone is listed in parentheses. \ The number of configurations eliminated by
combinations of rules may be difficult to tally because the rules interact in
complex ways. \ Together, these $8$ binary rules reduce the number of index
configurations from $16,384$ to $2,712$.

\bigskip

\emph{Rule 2.1}. $I_{\alpha,\beta}=-1/2$ $\&$ $\Lambda_{\alpha,\beta}%
=1$.$\,(4096)$

\emph{Rule 2.2}. $I_{\gamma}=-1$ $\&$ $\Lambda_{\gamma}=0$.$\,(4096)$.

\emph{Rule 2.3}. $\left\vert \tau_{\alpha,\beta}\right\vert =3/2$ $\&$
$\Lambda_{\alpha,\beta}=1$.$\,(4096)$

\emph{Rule 2.4 }$\left\vert \pi_{\alpha,\beta}\right\vert =3/2$ $\&$
$\Lambda_{\alpha,\beta}=1$.$\,\ (4096)$

\emph{Rule 2.5 }$\left\vert \tau_{\alpha,\beta}\right\vert =3/2$ $\&$
$d\tau_{\alpha,\beta}\neq\tau_{\alpha,\beta}$.$\,\ (6144)$

\emph{Rule 2.6 }$\left\vert \pi_{\alpha,\beta}\right\vert =3/2$ $\&$
$d\pi_{\alpha,\beta}\neq\pi_{\alpha,\beta}$.$\,(6144)$.

\emph{Rule 2.7 }$\left\vert \tau_{\alpha,\beta}\right\vert =3/2$ $\&$
sign$\left(  d\tau_{\gamma}\right)  \neq$ sign$\left(  \tau_{\alpha,\beta
}\right)  $.$\,(4096)$

\emph{Rule 2.8 }$\left\vert \pi_{\alpha,\beta}\right\vert =3/2$ $\&$
sign$\left(  d\pi_{\gamma}\right)  \neq$ sign$\left(  \pi_{\alpha,\beta
}\right)  $.$\,(4096)$.

\subsubsection{Ternary Rules}

The five forbidden ternary selection rules listed below acting together
(without the binary rules) reduce the number of index configurations from
$16,384$ to $8,064$.

\bigskip

\emph{Rule 3.1}. $I_{\alpha,\beta}=1/2$ $\&$ $\left\vert \tau_{\alpha,\beta
}\right\vert =3/2$ $\&\ \left\vert \pi_{\alpha,\beta}\right\vert
=3/2$.$\,(2048)$

\emph{Rule 3.2}. $I_{\alpha,\beta}=1/2$ $\&$ $\Lambda_{\gamma}=0$ $\&$
$\left\vert \tau_{\alpha,\beta}\right\vert =3/2$.$\,(2048)$

\emph{Rule 3.3}. $I_{_{\alpha,\beta}}=-1/2$ $\&$ $\Lambda_{\gamma}=0$ $\&$
$\left\vert \tau_{\alpha,\beta}\right\vert =1/2$.$\,(2048)$

\emph{Rule 3.4}. $I_{\gamma}=-1$ $\&$ $\left\vert \tau_{\alpha,\beta
}\right\vert =3/2$ $\&\ \left\vert \pi_{\alpha,\beta}\right\vert
=3/2$.$\,(2048)$

\emph{Rule 3.5}. $\Lambda_{\gamma}=0$ $\&$ sign$\left(  d\tau_{\gamma}\right)
\neq$ sign$\left(  \tau_{\alpha,\beta}\right)  $.$\,(2048)$

\bigskip

\hspace*{-0.2in}These five rules cover all configurations not observed in the
$62$ 3D plots discussed above.

Except for Rules 2.1$-$2.4, which are based on known C point results or easily
verified geometrical constraints, the above rules are provisional, and await
confirmation from analysis. \ These rules, however, are $\emph{practical}$
rules, since any configuration not seen in $10^{6}$ samples in unlikely to
appear in an experiment. \ The fact that the rules reduce the number of
configurations only to $2104$, which is nearly twice the number actually
observed, Fig. \ref{Fig21}, may imply the existence of additional rules or may
be due to other, presently unknown, factors.

\subsection{Probabilities}

Listed below are the probabilities of occurrence of various configurations in
our database. \ We emphasize that these are not densities, and therefore do
not directly correspond to the quantities that would be measured in an
experiment. \ In order to compute densities each realization needs to be
weighted by a Jacobian (with units of $area^{-1}$) that, like the realization
itself, is a function of the wave field parameters in Eq. (\ref{GenFieldXpan}).

At present there are two Jacobians available [$9$]. \ Defining $\boldsymbol{J}%
=\boldsymbol{\nabla}\left(  \operatorname{Re}\left(  \boldsymbol{E\cdot
E}\right)  \right)  \times\boldsymbol{\nabla}\left(  \operatorname{Im}\left(
\boldsymbol{E\cdot E}\right)  \right)  $ these are
\end{subequations}
\begin{subequations}
\label{Jac23}%
\begin{align}
J_{2D}  &  =\left\vert \boldsymbol{J}_{z}\right\vert ,\label{J2D}\\
J_{3D}  &  =\left\vert \boldsymbol{J}\right\vert . \label{J3D}%
\end{align}
In Eq. (\ref{J2D}) the $z$-axis is normal to the plane of observation, here
$\Sigma$. \ $J_{2D}$ is suitable for counting 2D densities in the plane, such
as the density of C points that appear where C lines pierce the plane.
\ $J_{3D}$ measures 3D quantities, such as the length of C line/unit volume.
\ But neither of these two Jacobians, at least as written, depends
\emph{sensitively} on the angle between the $\Sigma$ and $\Sigma_{0}$. \ This
sensitivity is essential, because as emphasized above, the statistics change
importantly when $\Sigma$ rotates away from $\Sigma_{0}$ by a small, but
finite angle.

The probabilities listed below, although not directly connected to experiment,
are, in principle, calculable from theory. \ We note that eliminating the
Jacobian from the theory, in fact, greatly simplifies the calculation.

Among the endless probabilities $P$ that can be extracted from our database we
believe the following to be of greatest interest:

\bigskip

(1) Positive and negative values for $\tau_{\alpha,\beta}$ and $\pi
_{\alpha,\beta}$ occur with equal probability $P$, and$\ P\left(  \tau
_{\alpha,\beta}=\pm\frac{1}{2}\right)  =P\left(  \pi_{\alpha,\beta}=\pm
\frac{1}{2}\right)  =0.80$, $P\left(  \tau_{\alpha,\beta}=\pm\frac{3}%
{2}\right)  =P\left(  \pi_{\alpha,\beta}=\pm\frac{3}{2}\right)  =0.20$.
\ Throughout, we use the notation $\tau_{\alpha,\beta}=\pm\frac{1}{2}$,
$\pi_{\alpha,\beta}=\pm\frac{1}{2}$, instead of $\left\vert \tau_{\alpha
,\beta}\right\vert =\left\vert \pi_{\alpha,\beta}\right\vert =1/2$, etc., to
emphasize that in searching our database we search for each sign combination
separately. \ When in later tables the results are sign dependent we list each
sign combination.

\bigskip

(2) The frequency of occurrence of Mobius strips with one and three
half-twists that are lemons, monstars, and stars, are listed in Table 1. \ The
selection rules forbid lemons with $\tau_{\alpha,\beta}$ and $\pi
_{\alpha,\beta}$ equal to $\pm\frac{3}{2}$.

\bigskip
\end{subequations}
\begin{center}
Table 1. Lemon, monstar, star, and indices $\tau_{\alpha,\beta}$ and
$\pi_{\alpha,\beta}$.%

\begin{tabular}
[c]{|c|c|c|c|}\hline
$\tau_{\alpha,\beta},\pi_{\alpha,\beta},$ & Lemon ($I_{_{\alpha,\beta}}%
=\frac{1}{2},\Lambda_{_{\alpha,\beta}}=1)$ & Monstar ($I_{_{\alpha,\beta}%
}=\frac{1}{2},\Lambda_{_{\alpha,\beta}}=3)$ & Star ($I_{_{\alpha,\beta}%
}=-\frac{1}{2},\Lambda_{_{\alpha,\beta}}=3)$\\\hline
$\pm\frac{1}{2}$ & $0.39$ & $0.11$ & $0.30$\\\hline
$\pm\frac{3}{2}$ & $0$ & $0.005$ & $0.20$\\\hline
\end{tabular}

\end{center}

\bigskip

(3) Combinations of $\tau_{\alpha,\beta}$ and $\pi_{\alpha,\beta}$ are listed
in Table 2. \ As can be seen, it is some four times more likely that these two
indices are the same than that they differ.

\begin{center}
Table 2. Indices $\tau_{\alpha,\beta}$ and $\pi_{\alpha,\beta}$.%

\begin{tabular}
[c]{|c|r|r|}\hline
Indices & $\tau_{\alpha,\beta}=\pm\frac{1}{2}$ & $\pm\frac{3}{2}$\\\hline
\multicolumn{1}{|r|}{$\pi_{\alpha,\beta}=\pm\frac{1}{2}$} & $0.17$ &
$0.026$\\\hline
\multicolumn{1}{|r|}{$\pm\frac{3}{2}$} & $0.026$ & $0.025$\\\hline
\end{tabular}
\bigskip
\end{center}

(4) Table 3 lists probabilities for combinations of $\tau_{\alpha,\beta}$ and
$d\tau_{\alpha,\beta}$. \ These probabilities are the same for $\pi
_{\alpha,\beta}$ and $d\pi_{\alpha,\beta}$ combinations, and are sign
dependent. \ Combinations with zero probability are forbidden by selection
rules. \ For the case $\left\vert \tau_{\alpha,\beta}\right\vert $
$(\left\vert \pi_{\alpha,\beta}\right\vert )=\frac{1}{2}$, the number of
occurrences in which $d\tau_{\alpha,\beta}$ $(d\pi_{\alpha,\beta})\neq
\tau_{\alpha,\beta}$ $(\pi_{\alpha,\beta})$ exceeds those in which
$d\tau_{\alpha,\beta}$ $(d\pi_{\alpha,\beta})=\tau_{\alpha,\beta}$
$(\pi_{\alpha,\beta})$, reflecting the fact that somewhat more than half of
all M\"{o}bius strips have a complex structure containing an additional, self
cancelling pair of half-twists, (see Fig. \ref{Fig9}).

\begin{center}

Table 3. Indices $\tau_{\alpha,\beta}$ $(\pi_{\alpha,\beta})$ and
$d\tau_{\alpha,\beta}$ $(d\pi_{\alpha,\beta})$.%

\begin{tabular}
[c]{|c|r|r|r|r|}\hline
Indices & $\tau_{\alpha,\beta}$ $(\pi_{\alpha,\beta})=+\frac{1}{2}$ &
$-\frac{1}{2}$ & $+\frac{3}{2}$ & $-\frac{3}{2}$\\\hline
\multicolumn{1}{|r|}{$d\tau_{\alpha,\beta}$ $(d\pi_{\alpha,\beta})=+\frac
{1}{2}$} & $0.19$ & $0.11$ & $0$ & $0$\\\hline
\multicolumn{1}{|r|}{$-\frac{1}{2}$} & $0.11$ & $0.19$ & $0$ & $0$\\\hline
\multicolumn{1}{|r|}{$+\frac{3}{2}$} & $0.09$ & $0.008$ & $0.10$ & $0$\\\hline
\multicolumn{1}{|r|}{$-\frac{3}{2}$} & $0.008$ & $0.09$ & $0$ & $0.10$\\\hline
\end{tabular}

\bigskip
\end{center}

(5) Correlations between the orientations of axis $\boldsymbol{\gamma}$ and
axes $\boldsymbol{\alpha},\boldsymbol{\beta}$ are reflected in Tables 4 and 5.
\ Worthy of note is that the possibly surprising statistical equivalence of
$\tau_{\alpha,\beta}$, and $\pi_{\alpha,\beta}$ evident in all the other
tables is broken in Table 5 for $\Lambda_{\gamma}$.

\begin{center}

Table 4. Indices $I_{\gamma}\ $and $\tau_{\alpha,\beta}$ $(\pi_{\alpha,\beta
})$.%

\begin{tabular}
[c]{|c|r|r|}\hline
Indices & $\tau_{\alpha,\beta}$ $(\pi_{\alpha,\beta})=\pm\frac{1}{2}$ &
$\pm\frac{3}{2}$\\\hline
\multicolumn{1}{|r|}{$I_{\gamma}=+1$} & $0.16$ & $0.086$\\\hline
\multicolumn{1}{|r|}{$-1$} & $0.23$ & $0.014$\\\hline
\end{tabular}
\bigskip

\bigskip

Table 5. Indices $\Lambda_{\alpha,\beta}$, $\Lambda_{\gamma}$, $\tau
_{\alpha,\beta}$, and $\pi_{\alpha,\beta}$.%

\begin{tabular}
[c]{|c|r|r|r|r|}\hline
Indices & $\tau_{\alpha,\beta}$ $=\pm\frac{1}{2}$ & $\pm\frac{3}{2}$ &
$\pi_{\alpha,\beta}=\pm\frac{1}{2}$ & $\pm\frac{3}{2}$\\\hline
\multicolumn{1}{|r|}{$\Lambda_{\alpha,\beta}=1$} & $0.19$ & $0$ & $0.19$ &
$0$\\\hline
\multicolumn{1}{|r|}{$3$} & $0.21$ & $0.10$ & $0.21$ & $0.10$\\\hline
\multicolumn{1}{|r|}{$\Lambda_{\gamma}=0$} & $0.073$ & $0.073$ & $0.10$ &
$0.042$\\\hline
\multicolumn{1}{|r|}{$4$} & $0.33$ & $0.028$ & $0.30$ & $0.059$\\\hline
\end{tabular}

\end{center}

\bigskip

\section{SUMMARY}

Prior studies of C points and C lines have concentrated on a single projection
of the major and minor axes of the ellipses surrounding a C line onto a plane,
in most instances the plane we call $\Sigma_{0}$, the plane of the C circle.
\ Examining the full 3D arrangement of the surrounding ellipses we found that
their major and minor axes, $\boldsymbol{\alpha}$ and $\boldsymbol{\beta}$,
generate M\"{o}bius strips, whereas the normals to these ellipses,
$\boldsymbol{\gamma}$, generate a section of a cone. \ The M\"{o}bius strips
have either one or three half-twists, and form segments of either right-handed
or left-handed screws.

The $\Sigma_{0}$ projections of $\boldsymbol{\alpha}$ and $\boldsymbol{\beta}$
give rise to two well-known indices [$1,3$]: the conserved topological index
$I_{\alpha,\beta}$, and the highly useful, albeit nontopological, line
characterization $\Lambda_{\alpha,\beta}$. \ These, and all other indices in
$\Sigma_{0}$, have the same values for axes $\boldsymbol{\alpha}$ and
$\boldsymbol{\beta}$, whence the subscript $\alpha,\beta$.

$I_{\alpha,\beta}$ measures the rotation of the projections of
$\boldsymbol{\alpha},\boldsymbol{\beta}$ around the C point, and takes on the
generic values $I_{\alpha,\beta}=\pm1/2$. \ On a circular path centered on the
C point, here labelled $\sigma_{0}$, $\Lambda_{\alpha,\beta}$ counts the
number of times an axis projection points directly at the C point.
\ Generically, $\Lambda_{\alpha,\beta}=1,3$, and there are three possible
combinations of $I_{\alpha,\beta}$ and $\Lambda_{\alpha,\beta}$: the lemon
($I_{\alpha,\beta}=1/2,\Lambda_{\alpha,\beta}=1$); the monstar ($I_{\alpha
,\beta}=1/2,\Lambda_{\alpha,\beta}=3$); and the star ($I_{\alpha,\beta
}=-1/2,\Lambda_{\alpha,\beta}=3$).

We introduced $8$ new indices.

Two of these involve the projections of axis $\boldsymbol{\gamma}$ onto
$\Sigma_{0}$. \ As is the case for axes $\boldsymbol{\alpha}$ and
$\boldsymbol{\beta}$, for axis $\boldsymbol{\gamma}$ this projection gives
rise to a conserved topological index, $I_{\gamma}$, and a line
characterization, $\Lambda_{\gamma}$. \ Generically $I_{\gamma}=\pm1$, and
$\Lambda_{\gamma}=0,4$, and also here there are only three possible
combinations: $I_{\gamma}=1,\Lambda_{\gamma}=0,4$; $I_{\gamma}=-1,\Lambda
_{\gamma}=4$.

Six new indices arise from two new projections: the $\tau_{0}$ projection and
the $\pi_{0\text{ }}$projection.

The major and minor axes of the surrounding ellipses generate two winding
numbers for each projection: $\tau_{\alpha,\beta}$ and $d\tau_{\alpha,\beta}$
for the $\tau_{0}$ projection, $\pi_{\alpha,\beta}$ and $d\pi_{\alpha,\beta}$
for the $\pi_{0}$ projection. \ Each of these four indices can take on four
different values: $\pm1/2,\pm3/2$. \ $\tau_{\alpha,\beta}$ measures the number
of half-twists of the M\"{o}bius strip, the other three indices measure other,
more subtle structural features. \ $\tau_{\alpha,\beta}$ and $\pi
_{\alpha,\beta}$ are independent, and all $16$ combinations of these indices
are found. \ $\tau_{\alpha,\beta}$ and $d\tau_{\alpha,\beta}$, and also
$\pi_{\alpha,\beta}$ and $d\pi_{\alpha,\beta}$, are not completely
independent, and only $10$ of the $16$ possible combinations of each index
pair are allowed.

The $\tau_{0}$ and $\pi_{0\text{ }}$projections of axis $\boldsymbol{\gamma}$
generate the last two of our $8$ new indices. \ These are $d\tau_{\gamma}$ and
$d\pi_{\gamma}$, each of which can be $\pm2$.

A set of simple rules, the phase ratchet rules, were formulated that permit
$\tau_{\alpha,\beta}$ and $\pi_{\alpha,\beta}$ to be determined by inspection
from the $\tau_{0}$ and $\pi_{0\text{ }}$projections; $d\tau_{\alpha,\beta}$
and $d\pi_{\alpha,\beta}$ can also be determined by inspection from these projections.

Of the $16,384$ combinations of indices that would be possible in the absence
of all restrictions, some $1,150$ were observed in a database containing
$10^{6}$ randomly generated realizations. \ This number of distinct C lines
exceeds by a factor of $\sim350$ the three types of lines (lemon, monstar,
star) recognized previously. \ Thirteen selection rules were formulated that
reduce the number of possibilities to $2,104$. \ These rules include the
well-known restriction that gives rise to the lemon-monstar-star trio,
together with $12$ new rules; the most important of these new rules are: all
lemons are one-half-twist Mobius strips, and [$34$], all three-half-twist M\"{o}bius
strips are either stars or monstars. \ The converse of the lemon rule,
however, does not hold, and one-half-twist M\"{o}bius strips can also be
monstars or stars.

Statistics of the most important configurations were also presented. \ Most
notable of these is that $80\%$ of all M\"{o}bius strips have a single
half-twist, the remaining $20\%$ three half-twists. \ Also noteworthy is the
fact that although all three-half-twist strips are monstars or stars, the
majority of monstars and stars are, in fact, one-half-twist M\"{o}bius strips.

A recurrent theme was the complexity of the various M\"{o}bius strips; this
complexity was illustrated by numerous examples, and described geometrically.
\ Because of this complexity, a more complete mathematical description of
these objects, which includes the dependence of the various indices, and their
selection rules, on the parameters of the optical field, may be rather complicated.

Recent advances make experimental measurements of the M\"{o}bius strips and
cones feasible in both the microwave and optical regions of the spectrum.
\ Such experiments would do much to further our knowledge of the structure of
real, as opposed to simulated, 3D optical fields.

\bigskip
\begin{center}
{\large \textbf{Acknowledgements}}
\end{center}

I am pleased to acknowledge extensive discussion with, and many helpful
comments and suggestions by, Prof. David A. Kessler, and an important email
from Dr. Mark R. Dennis (Bristol).

\bigskip
\begin{center}
{\large \textbf{References}}
\end{center}

\hspace*{-0.23in}[1] J. F. Nye, \emph{Natural Focusing and Fine Structure of
Light} (IOP Publ., Bristol, 1999).

\hspace*{-0.23in}[2] J. F. Nye, \textquotedblleft Polarization effects in the
diffraction of electromagnetic waves: the role of
disclinations,\textquotedblright\ Proc. Roy. Soc. Lond. A \textbf{387},
105$-$132 (1983).

\hspace*{-0.23in}[3] J. F. Nye, \textquotedblleft Lines of circular
polarization in electromagnetic wave fields,\textquotedblright\ Proc. Roy.
Soc. Lond. A \textbf{389}, 279$-$290 (1983).

\hspace*{-0.23in}[4] J. F. Nye and J. V. Hajnal, \textquotedblleft The wave
structure of monochromatic electromagnetic radiation,\textquotedblright\ Proc.
Roy. Soc. Lond. A \textbf{409}, 21$-$36 (1987).

\hspace*{-0.23in}[5] J. V. Hajnal, \textquotedblleft Singularities in the
transverse fields of electromagnetic waves. I. Theory,\textquotedblright
\ Proc. Roy. Soc. Lond. A \textbf{414}, 433$-$446 (1987).

\hspace*{-0.23in}[6] J. V. Hajnal, \textquotedblleft Singularities in the
transverse fields of electromagnetic waves. II Observations on the electric
field,\textquotedblright\ Proc. Roy. Soc. Lond. A \textbf{414}, 447$-$468 (1987).

\hspace*{-0.23in}[7] J. V. Hajnal, \textquotedblleft Observations of
singularities in the electric and magnetic fields of freely propagating
microwaves,\textquotedblright\ Proc. Roy. Soc. Lond. A \textbf{430}, 413$-$421 (1990).

\hspace*{-0.23in}[8]. M. V. Berry,\textquotedblleft Geometry of phase and
polarization singularities, illustrated by edge diffraction and the
tides,\textquotedblright\ in Second International Conference on Singular
Optics, M. S. Soskin and M. V. Vasnetsov Eds., Proc. SPIE \textbf{4403,}
1$-$12 (2001).

\hspace*{-0.23in}[9] M. V. Berry and M. R. Dennis, \textquotedblleft
Polarization singularities in isotropic random vector waves,\textquotedblright\ Proc. Roy. Soc.
Lond. A \textbf{457}, 141$-$155 (2001).

\hspace*{-0.23in}[10] M. V. Berry, \textquotedblleft Index formulae for
singular lines of polarization,\textquotedblright\ J. Opt. A \textbf{6},
675$-$678 (2004).

\hspace*{-0.23in}[11] I. Freund, \textquotedblleft Polarization singularities
in optical lattices\textquotedblright\ Opt. Lett. \textbf{29}, 875$-$877\ (2004).

\hspace*{-0.23in}[12] I. Freund, \textquotedblleft Polarization singularity
proliferation in three-dimensional ellipse fields,\textquotedblright\ Opt.
Lett. \textbf{30}, 433$-$435 (2005).

\hspace*{-0.23in}[13] I. Freund, \textquotedblleft Polarization singularity
anarchy in three dimensional ellipse fields,\textquotedblright\ Opt. Commun.
\textbf{242}, 65$-$78 (2004).

\hspace*{-0.23in}[14] I. Freund, \textquotedblleft Cones, spirals, and Mobius
strips in elliptically polarized light,\textquotedblright\ Opt. Commun.
\textbf{249}, 7$-$22\ (2005).

\hspace*{-0.23in}[15] I. Freund, \textquotedblleft Hidden order in optical
ellipse fields: I. Ordinary ellipses,\textquotedblright\ Opt. Commun.
\textbf{256}, 220$-$241\ (2005).

\hspace*{-0.23in}[16] F. Flossmann, K. O`Holleran, M. R. Dennis, and M. J.
Padgett, \textquotedblleft Polarization Singularities in 2D and 3D Speckle
Fields,\textquotedblright\ Phys. Rev. Lett. \textbf{100}, 203902 (2008).

\hspace*{-0.23in}[17] M. Born and E. W. Wolf, \emph{Principles of Optics}
(Pergamon Press, Oxford, 1959).

\hspace*{-0.23in}[18] M. V. Berry and J. H. Hannay, \textquotedblleft Umbilic
points on Gaussian random surfaces,\textquotedblright\ J. Phys. A \textbf{10},
1809$-$1821 (1977).

\hspace*{-0.23in}[19] S. Zhang and A. Z. Genack, \textquotedblleft Statistics
of Diffusive and Localized Fields in the Vortex Core,\textquotedblright\ Phys.
Rev. Lett. \textbf{99}, 203901 (2007).

\hspace*{-0.23in}[20] S. Zhang, B. Hu, P. Sebbah, and A. Z. Genack,
\textquotedblleft Speckle Evolution of Diffusive and Localized
Waves,\textquotedblright\ Phys. Rev. Lett. \textbf{99}, 063902 (2007).

\hspace*{-0.23in}[21] R. Dandliker, I. Marki, M. Salt, and A. Nesci,
\textquotedblleft Measuring optical phase singularities at subwavelength
resolution,\textquotedblright\ J. Optics A \textbf{6}, S189$-$S196 (2004).

\hspace*{-0.23in}[22] P. Tortora, R. Dandliker, W. Nakagawa, and L. Vaccaro,
\textquotedblleft Detection of non-paraxial optical fields by optical fiber
tip probes,\textquotedblright\ Opt. Commun. \textbf{259}, 876$-$882 (2006).

\hspace*{-0.23in}[23] C. Rockstuhl, I. Marki, T. Scharf, M. Salt, H. P.
Herzig, and R. Dandliker, \textquotedblleft High resolution interference
microscopy: A tool for probing optical waves in the far-field on a nanometric
length scale,\textquotedblright\ Current Nanoscience \textbf{2}, 337$-$350 (2006).

\hspace*{-0.23in}[24] P. Tortora, E. Descrovi, L. Aeschimann, L. Vaccaro, H.
P. Herzig, and R. Dandliker, \textquotedblleft Selective coupling of HE11 and
TM01 modes into microfabricated fully metal-coated quartz
probes,\textquotedblright\ Ultramicroscopy \textbf{107}, 158$-$165 (2007).

\hspace*{-0.23in}[25] K. G. Lee, H. W. Kihm, J. E. Kihm, W. J. Choi, H. Kim,
C. Ropers, D. J. Park, Y. C. Yoon, S. B. Choi, H. Woo, J. Kim, B. Lee, Q. H.
Park, C. Lienau C, and D. S. Kim, \textquotedblleft Vector field microscopic
imaging of light,\textquotedblright\ Nature Photonucs \textbf{1}, 53$-$56 (2007).

\hspace*{-0.23in}[26] Z. H. Kim and S. R. Leone, \textquotedblleft
Polarization-selective mapping of near-field intensity and phase around gold
nanoparticles using apertureless near-field microscopy,\textquotedblright%
\ Opt. Express \textbf{16}, 1733$-$1741 (2008).

\hspace*{-0.23in}[27] M. Burresi, R. J. Engelen, A. Opheij, D. van Oosten, D.
Mori, T. Baba, and L. Kuipers, \textquotedblleft Observation of Polarization
Singularities at the Nanoscale,\textquotedblright\ Phys. Rev. Lett.
\textbf{102}, 033902 (2009).

\hspace*{-0.23in}[28] R. J. Engelen, D. Mori, T. Baba, and L. Kuipers,
\textquotedblleft Subwavelength Structure of the Evanescent Field of an
Optical Bloch Wave,\textquotedblright\ Phys. Rev. Lett. \textbf{102}, 023902
(2009);Erratum: ibid. 049904 (2009).

\hspace*{-0.23in}[29] I. Freund, \textquotedblleft Coherency matrix
description of optical polarization singularities,' J. Opt. A \textbf{6},
S229$-$S234 (2004).

\hspace*{-0.23in}[30] M. R. Dennis, \textquotedblleft Geometric interpretation
of the 3-dimensional coherence matrix for nonparaxial
polarization,\textquotedblright\ J. of Opt. A \textbf{6}, 26$-$31 (2004).

\hspace*{-0.23in}[31] I. Freund, \textquotedblleft Critical point explosions
in two-dimensional wave fields,\textquotedblright\ Opt. Commun. \textbf{159},
99$-$112 (1999); ibid. \textbf{173}, 435 (2000).

\hspace*{-0.23in}[32] I. Freund, \textquotedblleft Polarization singularity
indices in Gaussian laser beams,\textquotedblright\ Opt. Commun. \textbf{201},
251$-$270 (2002).

\hspace*{-0.23in}[33] M. R. Dennis, \textquotedblleft Polarization
singularities in paraxial vector fields: morphology and
statistics,\textquotedblright\ Opt. Commun. \textbf{213}, 201$-$221 (2002).

\hspace*{-0.23in}[34] Dr. Mark R. Dennis (Bristol), private communication.

\end{document}